\documentclass[aps,prc,groupedaddress,twocolumn,nofootinbib]{revtex4-2}

\usepackage{bm}
\usepackage{graphicx}
\usepackage{ascmac}
\usepackage{amsmath,amssymb}
\usepackage{cancel}
\usepackage{braket}

\begin{document}


\title{Compositeness of $T_{cc}$ and $X(3872)$ by considering decay and coupled-channels effects}


\author{Tomona Kinugawa}
\email[]{kinugawa-tomona@ed.tmu.ac.jp}
\affiliation{Department of Physics, Tokyo Metropolitan University, Hachioji 192-0397, Japan}
\author{Tetsuo Hyodo}
\email[]{hyodo@tmu.ac.jp}
\affiliation{Department of Physics, Tokyo Metropolitan University, Hachioji 192-0397, Japan}

\date{\today}

\begin{abstract}
The compositeness of weakly bound states is discussed using the effective field theory from the viewpoint of the low-energy universality. We introduce a model with coupling of the single-channel scattering to the bare state, and study the compositeness of the bound state by varying the bare state energy. In contrast to the naive expectation that the near-threshold states are dominated by the molecular structure, we demonstrate that a non-composite state can always be realized even with a small binding energy. At the same time, however, it is shown that a fine tuning is necessary to obtain the non-composite weakly bound state. In other words, the probability of finding a model with the composite dominant state becomes larger with the decrease of the binding energy in accordance with the low-energy universality. For the application to exotic hadrons, we then discuss the modification of the compositeness due to the decay and coupled-channels effects. We quantitatively show that these contributions suppress the compositeness, because of the increase of the fraction of other components. Finally, as examples of near-threshold exotic hadrons, the structures of $T_{cc}$ and $X(3872)$ are studied by evaluating the compositeness. We find the importance of the coupled-channels and decay contributions for the structures of $T_{cc}$ and $X(3872)$, respectively. 

\end{abstract}

\maketitle


\section{Introduction}
\label{sec:intro}

Clarifying the internal structure of exotic hadrons is one of the central aims of  hadron physics. 
The recent observations of the exotic hadron candidates in the heavy hadron sectors provide opportunities for intensive studies on the structure of hadrons~\cite{Guo:2017jvc,Brambilla:2019esw}. Exotic hadrons are considered to have different internal structures from ordinary hadrons with $qqq$ or $q\bar{q}$ as described in the quark models. 
 
It is remarkable that many exotic hadron candidates have been discovered near two-hadron thresholds.  For example, the tetraquark $T_{cc}$ was observed slightly below the threshold of $D^{0}D^{*+}$ in the $T_{cc}\to D^{0}D^{0}\pi^{+}$ decay by the LHCb Collaboration in 2021~\cite{LHCb:2021auc,LHCb:2021vvq}. Its minimum quark content $cc\bar{u}\bar{d}$ indicates that $T_{cc}$ is a genuine exotic state with charm $C=+2$. As a charmonium-like state with $C=0$, $X(3872)$ was observed near the $D^{0}\bar{D}^{*0}$ threshold in the $B^{\pm}\to K^{\pm}\pi^{+}\pi^{-}J/\psi$ decay in 2003 by the Belle Collaboration~\cite{Belle:2003nnu}. $X(3872)$ is considered to be exotic because its mass is not in accordance with the corresponding energy predicted by the quark model~\cite{Godfrey:1985xj}. 

As possible internal structures of the exotic hadrons, hadronic molecule states and multiquark states are among those considered. The hadronic molecule state is a loosely bound composite system of hadrons formed by hadronic interactions; an example is the deuteron. In contrast, the multiquark state is a compact state of at least four quarks. To reveal the internal structure of the exotic hadrons, many studies are being performed from both the theoretical and experimental sides. 
 
The molecular nature of the bound state can be quantitatively studied by using the compositeness~\cite{Weinberg:1962hj,Weinberg:1963zza,Weinberg:1964zza,Weinberg:1965zz,Hyodo:2011qc,Aceti:2012dd,Hyodo:2013nka,Aceti:2014ala}. The compositeness is defined as the probability of finding the hadronic molecule component in the bound state. Theoretically, the compositeness can be evaluated either from the weak-binding relation~\cite{Weinberg:1965zz,Kamiya:2015aea,Kamiya:2016oao,Kinugawa:2022fzn} or from the residue of the pole of the scattering amplitude~\cite{Hyodo:2011qc,Aceti:2012dd}. The internal structure of many hadrons has been studied using the compositeness~\cite{Weinberg:1965zz,Baru:2003qq,Xiao:2012vv,Hyodo:2013iga,Sekihara:2013sma,Chen:2013upa,Aceti:2014ala,Aceti:2014wka,MartinezTorres:2014kpc,Sekihara:2014kya,Sekihara:2014qxa,Navarra:2015iea,Garcia-Recio:2015jsa,Meissner:2015mza,Kamiya:2015aea,Sekihara:2015gvw,Guo:2015daa,Guo:2016wpy,Lu:2016gev,Kamiya:2016oao,Kang:2016jxw,Esposito:2021vhu,Li:2021cue,Kinugawa:2022fzn,Song:2022yvz,Albaladejo:2022sux,Mikhasenko:2022rrl}. The notion of compositeness has also been applied to other systems, such as nuclei and atoms~\cite{Braaten:2003sw,Duine:2003zz,Schmidt:2011zu,Kinugawa:2022fzn}.

The phenomena associated with the near-threshold states are governed by the low-energy universality~\cite{Braaten:2004rn,Naidon:2016dpf}. 
From the universality argument, it is expected that the near-threshold states are dominated by the molecular component~\cite{Hanhart:2014ssa,Kinugawa:2022fzn}. In fact, Ref.~\cite{Hyodo:2014bda} shows that the $s$-wave bound states become completely composite in the weak-binding limit. A similar discussion has been given regarding the cluster phenomena in nuclear physics, such as the ground state of ${}^{8}$Be and the ${}^{12}$C Hoyle state~\cite{PTPS52.89}. From these discussions of the near-threshold states, one may naively expect that $T_{cc}$ and $X(3872)$ are the composite dominant states.

\begin{figure}
\centering
\includegraphics[width=0.45\textwidth]{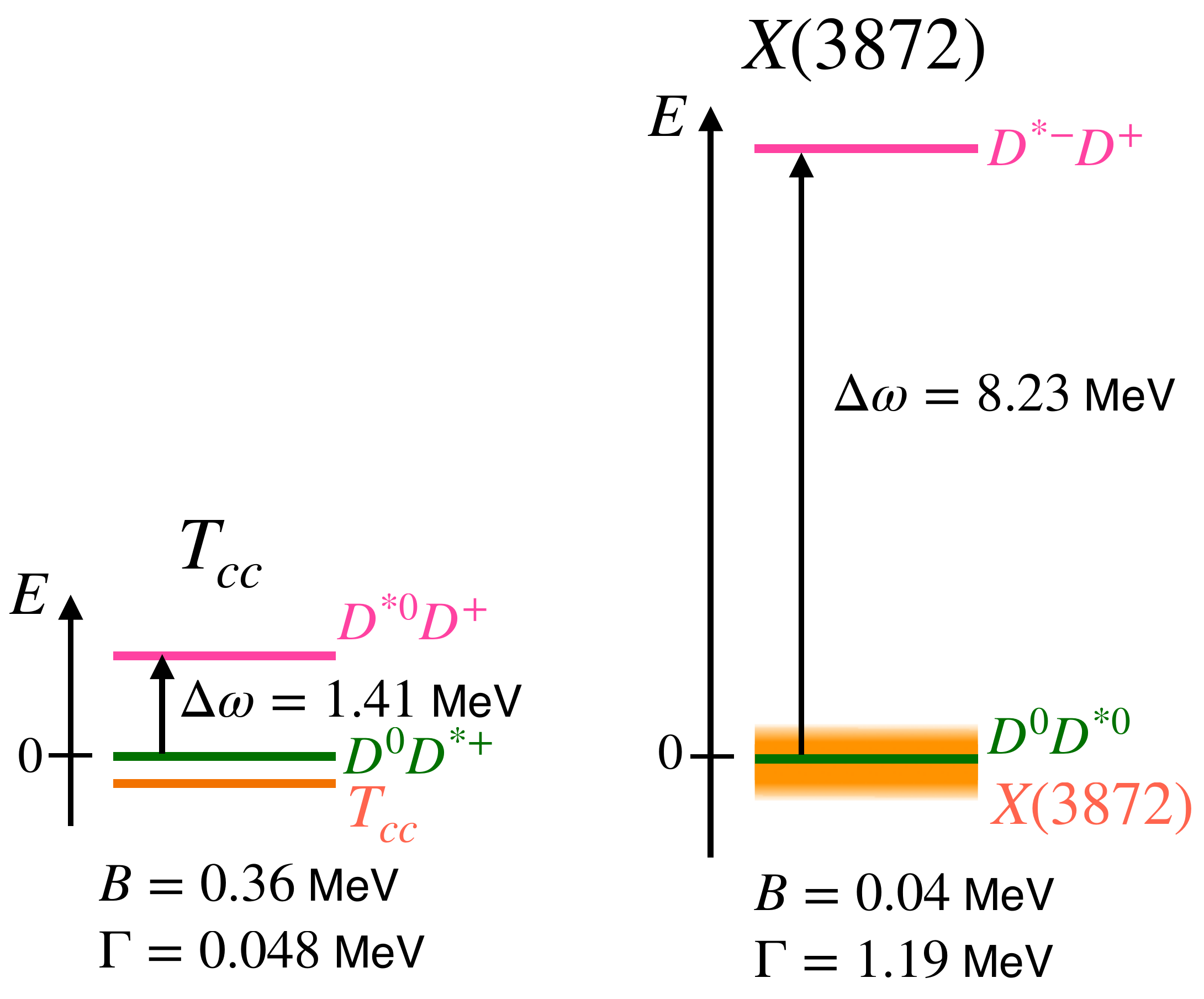}
\caption{Schematic illustrations of the $T_{cc}$ system (left) and $X(3872)$ system (right).}
\label{fig:Tcc-X3872}
\end{figure}

However, the small binding energy is not the only characteristic feature of $T_{cc}$ and $X(3872)$. First, both $T_{cc}$ and $X(3872)$ decay strongly and have a finite decay width. Next, the threshold channel [$D^{0}D^{*+}$ for $T_{cc}$ and $D^{0}\bar{D}^{*0}$ for $X(3872)$] has an isospin partner [$D^{*0}D^{+}$ for $T_{cc}$ and $D^{*-}D^{+}$ for $X(3872)$] at a slightly higher energy. These features are illustrated in Fig.~\ref{fig:Tcc-X3872}. It is shown that these decay and coupled-channels contributions modify the compositeness of the bound state~\cite{Kamiya:2016oao}. To understand the nature of $T_{cc}$ and $X(3872)$, we need to quantitatively evaluate the contributions from the decay and channel coupling to the compositeness.

In this work, we first demonstrate how the expectation of the molecular nature of the near-threshold states is realized in an explicit model calculation. We show that the shallow bound state can be elementary dominant only with a fine tuning of the model parameter. In most of the parameter region except for the fine-tuned case, the weakly bound state is composite dominant, as expected from the universality. To consider the realistic exotic hadrons, we then examine the effects of the decay and the coupled channel to the compositeness. We quantitatively evaluate the modification of the expectation from the universality due to the decay and the coupled-channels effects. We finally apply the model to calculate the compositeness of $T_{cc}$ and $X(3872)$ to clarify the important effect for these states. 

This paper is organized as follows. In Sec.~\ref{sec:wbs}, we introduce the effective field theory, and numerically calculate the compositeness to discuss the nature of the shallow bound state. We then consider the contributions of the four-point contact interaction, the decay, and the channel coupling to the compositeness in Sec.~\ref{sec:effects}. In Sec.~\ref{sec:apply}, we estimate the compositeness of $T_{cc}$ and $X(3872)$ by focusing on the importance of the decay and the coupled-channels effects. A summary of this work is given in Sec. V.


\section{Weakly bound states and low-energy universality}
\label{sec:wbs}
In this section, we discuss the composite nature of the weakly  bound states in relation to the low-energy universality. In Sec.~\ref{subsec:eft-2}, we first construct a simple scattering model with the nonrelativistic effective field theory where a bound state originates from the bare state. In Sec.~\ref{subsec:calc-2}, we then numerically compare the compositeness of typical and weakly bound states, and discuss the deviation from the expectation of the low-energy universality with a finite binding energy. We also examine the validity of the weak-binding relation in this model in Sec.~\ref{subsec:wbr}.

\subsection{Effective field theory}
\label{subsec:eft-2}

Let us introduce a nonrelativistic effective field theory to consider the compositeness of the bound state. We construct a model which describes the single-channel scattering of $\psi_{1}$ and $\psi_{2}$ coupled to the discrete state $\phi$ without the direct $\psi_{1}$ and $\psi_{2}$ interactions. The Hamiltonian is
\begin{align}
\mathcal{H}_{\rm free}&=\frac{1}{2m_{1}}{\nabla}\psi_{1}^{\dag}\cdot {\nabla}\psi_{1}+\frac{1}{2m_{2}}{\nabla}\psi_{2}^{\dag}\cdot {\nabla}\psi_{2}
\nonumber \\
&\quad +\frac{1}{2M}{\nabla}\phi^{\dag}\cdot {\nabla}\phi+\nu_{0}\phi^{\dag}\phi, 
\label{eq:H-free-2}\\
 \mathcal{H}_{\rm int}&=g_{0}(\phi^{\dagger}\psi_{1}\psi_{2}+\psi^{\dagger}_{1}\psi^{\dagger}_{2}\phi).
\label{eq:H-wbs}
\end{align}
Here $m_{1}$, $m_{2}$, and $M$ are the masses of $\psi_{1}$, $\psi_{2}$, and the discrete (bare) state $\phi$, respectively. $\nu_{0}$ is the energy of the bare state $\phi$ measured from the $\psi_{1}\psi_{2}$ threshold, and $g_{0}$ is the bare coupling constant of the contact three-point interaction. For the Hamiltonian in Eq.~\eqref{eq:H-wbs} to be Hermitian, $g_{0}$ must be real. This model can also be regarded as the resonance model without the direct $\psi_{1}\psi_{2}$ interaction in Refs.~\cite{Braaten:2007nq,Kinugawa:2022fzn}.

In this paper, we focus on the two-body scattering of $\psi_{1}$ and $\psi_{2}$. While we have no direct interactions in this model, the $\psi_{1}\psi_{2}$ scattering occurs through the intermediate $\phi$ state. Regarding the s-channel exchange of $\phi$ as the effective interaction $V(k)$, we can derive the on-shell T-matrix $T_{\rm on}(k)$ of the $\psi_{1}\psi_{2}$ scattering as a function of the on-shell momentum $k$ from the Lippmann-Schwinger equation:
\begin{align}
T_{\rm on}(k)&=V(k)+V(k)G(k)T_{\rm on}(k),
\label{eq:LS-eq} \\
V(k)&=\frac{g_{0}^{2}}{\frac{k^{2}}{2\mu}-\nu_{0}},
\label{eq:V-1ch} \\
G(k)&=\int\frac{d^{3}q}{(2\pi)^{3}}\frac{1}{\frac{k^{2}}{2\mu}-\frac{q^{2}}{2\mu}+i0^{+}},
\label{eq:G-int}
\end{align}
with the reduced mass of the $\psi_{1}\psi_{2}$ system $\mu=(1/m_{1}+1/m_{2})^{-1}$. This model has no crossing symmetry because it is nonrelativistic. Therefore, there are no crossed-channel exchanges which are not realized by the vertices in the Hamiltonian. In fact, due to the particle number conservations, the on-shell T-matrix in Eq.~\eqref{eq:LS-eq} is exact in the two-body sector, as discussed in Refs.~\cite{Braaten:2007nq,Kamiya:2016oao}. Because $V(k)$ does not depend on the off-shell momenta, the Lippmann-Schwinger equation reduces to an algebraic equation. At the same time, the absence of the angular dependence of the interaction $V(k)$ leads to the $s$-wave scattering amplitude. To avoid the divergence of the $q$ integration in the loop function $G(k)$, a cutoff $\Lambda$ is introduced as the upper applicable boundary of the momentum in the effective field theory~\cite{Braaten:2007nq}. In this case, regularized $G(k)$ becomes
\begin{align}
G(k)&=-\frac{\mu}{\pi^{2}}\left[\Lambda+ik\arctan\left(-\frac{\Lambda}{ik}\right)\right].
\label{eq:G-1ch}
\end{align}

The scattering observables are expressed by the scattering amplitude $f(k)$, which is related with the on-shell T-matrix $T_{\rm on}(k)$ as $f(k)=-{\mu}/{(2\pi)}T_{\rm on}(k)$. From Eq.~\eqref{eq:LS-eq}, we obtain the scattering amplitude $f(k)$ as
\begin{align}
f(k)&=-\frac{\mu}{2\pi}\left[\frac{\frac{k^{2}}{2\mu}-\nu_{0}}{g_{0}^{2}}+\frac{\mu}{\pi^{2}}\left[\Lambda+ik\arctan\left(-\frac{\Lambda}{ik}\right)\right]\right]^{-1}.
\label{eq:1ch-amplitude}
\end{align}
For the low-energy scatterings, the inverse of the scattering amplitude $1/f(k)$ is expanded in powers of the momentum $k$ (the effective range expansion):
\begin{align}
\frac{1}{f(k)}&=-\frac{1}{a_{0}}+\frac{r_{e}}{2}k^{2}+\mathcal{O}(k^{4})-ik.
\label{eq:ERE}
\end{align}
The scattering length $a_{0}$ and the effective range $r_{e}$ are defined from the coefficients of the $k^{0}$ and $k^{2}$ terms in this expansion. By comparing Eq.~\eqref{eq:ERE} with the scattering amplitude in Eq.~\eqref{eq:1ch-amplitude}, we obtain the scattering length $a_{0}$ and the effective range $r_{e}$ in this model:
\begin{align}
a_{0}&=-\left[\frac{2\pi\nu_{0}}{g_{0}^{2}\mu}-\frac{2}{\pi}\Lambda\right]^{-1},
\label{eq:1ch-a}\\
r_{e}&=-\frac{2\pi}{g_{0}^{2}\mu^{2}}+\frac{4}{\pi\Lambda}.
\label{eq:1ch-re}
\end{align}
We note that the effective range in Eq.~\eqref{eq:1ch-re} has an upper bound $r_{e}\leq 4/(\pi\Lambda)$ because $g_{0}^{2}\geq 0$ for real $g_{0}$. For a finite cutoff $\Lambda$, the contact interaction in the Lagrangian is smeared to have a finite interaction range of the order of $\sim 1/\Lambda$. In this sense, the upper bound $4/(\pi\Lambda)$ can be regarded as the Wigner bound~\cite{Matuschek:2020gqe}, the upper bound of the effective range for finite range interactions. In the $\Lambda\to \infty$ limit under an appropriate renormalization, we obtain $r_{e}\leq 0$. This is a feature of the resonance model in the zero-range limit, as mentioned in Refs.~\cite{Braaten:2007nq,Kinugawa:2022fzn}. 

Suppose that this model generates a bound state with the binding energy $B$. The bound state is expressed as the pole of the scattering amplitude, and therefore the eigenmomentum is obtained by solving the bound state condition $f^{-1}(k)=0$. The bound state pole appears in the complex momentum plane at $k=i\kappa$ with $\kappa=\sqrt{2\mu B}>0$. The composite nature of the bound state can be characterized by the compositeness $X$~\cite{Weinberg:1962hj,Weinberg:1963zza,Weinberg:1964zza,Weinberg:1965zz,Hyodo:2011qc,Aceti:2012dd,Hyodo:2013nka,Aceti:2014ala} defined as the weight of the scattering states in the bound state $\ket{\Phi}$:
\begin{align}
X&=\int \frac{d^{3}k}{(2\pi)^{3}}|\braket{\bm{k}|\Phi}|^{2},
\label{eq:def-X}
\end{align}
where $\ket{\bm{k}}$ is the scattering eigenstate of the free Hamiltonian in Eq.~\eqref{eq:H-free-2} with the momentum $\bm{k}$. The compositeness $X$ can be expressed by the effective interaction and the loop function as discussed in Ref.~\cite{Kamiya:2016oao}:
\begin{align}
X&=\frac{G'(-B,\Lambda)}{G'(-B,\Lambda)-[V^{-1}(-B)]'}.
\label{eq:X-1ch-def}
\end{align}
Using this expression, we obtain $X$ in this model from Eqs.~\eqref{eq:V-1ch} and \eqref{eq:G-1ch}:
\begin{align}
X&=\left[1+\frac{\pi^{2}\kappa}{g_{0}^{2}\mu^{2}}\left(\arctan\left(\frac{\Lambda}{\kappa}\right)-\frac{\frac{\Lambda}{\kappa}}{1+\left(\frac{\Lambda}{\kappa}\right)^{2}}\right)^{-1}\right]^{-1}.
\label{eq:X-1ch}
\end{align}
We define the elementarity $Z$ as the overlap of the bound state $\ket{\Phi}$ with the bare state $\ket{\phi}$ which is the discrete eigenstate of the free Hamiltonian created by the bare $\phi$ field at rest:
\begin{align}
Z=|\braket{\phi|\Phi}|^{2}=\frac{-[V^{-1}(-B)]'}{G'(-B,\Lambda)-[V^{-1}(-B)]'}.
\label{eq:def-Z}
\end{align}
Namely, the elementarity $Z$ represents the fraction of the bare state component in the bound state. From the completeness relation with $\ket{\psi}$ and $\ket{\bm{k}}$, we obtain $Z+X=1$, which can also be directly seen from  Eqs.~\eqref{eq:X-1ch-def} and \eqref{eq:def-Z}.

Here we mention the model dependence of the compositeness $X$ in the system with the finite interaction range which corresponds to the finite cutoff $\Lambda$. In the $\Lambda\to \infty$ limit, the compositeness $X$ can be written using the observables such as the scattering length $a_{0}$, effective range $r_{e}$, and the radius of the bound state $R=1/\sqrt{2\mu B}$~\cite{Weinberg:1965zz,Kamiya:2016oao,Oller:2017alp,Kinugawa:2022fzn}:
\begin{align}
X&=\frac{a_{0}}{2R-a_{0}} = \frac{1}{1-r_{e}/R} = \sqrt{\frac{1}{1-2r_{e}/a_{0}}} .
\label{eq:wbs}
\end{align}
However, this expression in the zero-range limit does not work quantitatively when applied to the hadrons. In fact, the formula~\eqref{eq:wbs} gives $X\approx 1.69$ for the deuteron, when the experimental values of $a_{0},r_{e}$, and $R$ are substituted. Because the compositeness is defined as $0\leq X\leq 1$ for the stable bound state, this indicates the insufficiency of the formula~\eqref{eq:wbs} for the hadrons. This problem was recently discussed in Refs.~\cite{Kinugawa:2022fzn,Li:2021cue,Song:2022yvz,Albaladejo:2022sux}, and it has been shown that the finite range corrections are important to address this problem. 

To examine the effect of the finite interaction range, in this work, we keep $\Lambda$ finite in the effective field theory. The compositeness $X$ then depends on the value of the cutoff (interaction range) and on the regularization method (momentum dependence of the form factor), even for a given set of the experimental data. This is the ``model dependence'' of the compositeness we will study in the following, by varying the parameter $\nu_{0}$ in the effective field theory model. It should be noted that the model dependence becomes weakened and $X$ approaches the universal result~\eqref{eq:wbs} when the binding energy is small, as we will show in Sec.~\ref{subsec:wbr}. 

For later convenience, here we introduce the typical energy scale $E_{\rm typ}$ associated with the model. Because the cutoff $\Lambda$ gives the momentum scale, we define $E_{\rm typ}$ as
\begin{align}
E_{\rm typ}=\frac{\Lambda^{2}}{2\mu}.
\label{eq:Etyp}
\end{align}
If there is a bound state, the typical binding energy is expected to be $B\sim E_{\rm typ}$ based on the idea of naturalness~\cite{tHooft:1979rat,Veltman:1980mj,Kang:2016zmv,vanKolck:2022lqz}. We call the state with $B\ll E_{\rm typ}$ a weakly bound state. To ensure that the bound state is in the applicable region of the model, we impose the condition $B \leq E_{\rm typ}$.

\subsection{Numerical calculation}
\label{subsec:calc-2}

In this section, we numerically investigate the compositeness of bound states in the model given in Sec.~\ref{subsec:eft-2}. Before the concrete calculations, we summarize the relations among the model parameters. In principle, the model parameters---the bare state energy $\nu_{0}$, the coupling constant $g_{0}$, and the cutoff $\Lambda$---can be arbitrarily chosen. However, for a given binding energy $B$, the bound state condition $f^{-1}(i\kappa)=0$ leads to the expression of $g_{0}^{2}$ with other two parameters:
\begin{align}
g_{0}^{2}(B;\nu_{0},\Lambda)&=\frac{\pi^{2}}{\mu}(B+\nu_{0})\left[\Lambda-\kappa\arctan\left(\frac{\Lambda}{\kappa}\right)\right]^{-1},
\label{eq:g02}
\end{align}
with $\kappa=\sqrt{2\mu B}$. Therefore, we can reduce one degree of freedom by fixing the binding energy $B$. In addition, when we work with dimensionless quantities using $\Lambda$, the result does not depend on the specific value of $\Lambda$.

The remaining dimensionless parameter $\nu_{0}/E_{\rm typ}$ cannot be determined in the framework of the effective field theory.\footnote{$\nu_{0}$ is the energy of the discrete bare state and corresponds to the quark core state in the application to hadrons. The value of $\nu_{0}$ may be estimated, for instance, by the constituent quark model. } In this work, we vary $\nu_{0}/E_{\rm typ}$ within the allowed region to investigate the model dependence of the compositeness. The parameter region of $\nu_{0}/E_{\rm typ}$ is restricted as follows: (i) As we discussed below Eq.~\eqref{eq:Etyp}, the bound state should satisfy the condition $\kappa\leq \Lambda$, which leads to $\Lambda-\kappa\arctan(\Lambda/\kappa)>0$. Therefore, the sign of $g_{0}^{2}$ in Eq.~\eqref{eq:g02} coincides with the sign of $B+\nu_{0}$. The coupling constant square $g_{0}^{2}$ should be positive in Eq.~\eqref{eq:g02} for the Hermitian Hamiltonian. Hence the lower boundary of $\nu_{0}/E_{\rm typ}$ is given by $-B/E_{\rm typ}\leq\nu_{0}/E_{\rm typ}$.\footnote{Strictly speaking, the point $\nu_{0}=-B$ should be discussed with special care, because the coupling constant vanishes and hence the bound state pole decouples from the scattering amplitude~\cite{Hyodo:2014bda}. It is shown that the compositeness behaves as $X\to 1$ ($X\to 0$) with $g_{0}\to 0$ for fixed $B=0$ ($B\neq 0$). This behavior is confirmed in the present model as we will show below.} (ii) Because the effective field theory is applicable up to the energy scale $E_{\rm typ}$, the upper boundary of $\nu_{0}/E_{\rm typ}$ is given by $E_{\rm typ}/E_{\rm typ}=1$. In summary, the allowed $\nu_{0}/E_{\rm typ}$ region is determined as 
\begin{align}
-B/E_{\rm typ}\leq \nu_{0}/E_{\rm typ}\leq 1.
\label{eq:nu0-region}
\end{align}

\begin{figure}
\centering
\includegraphics[width=0.5\textwidth]{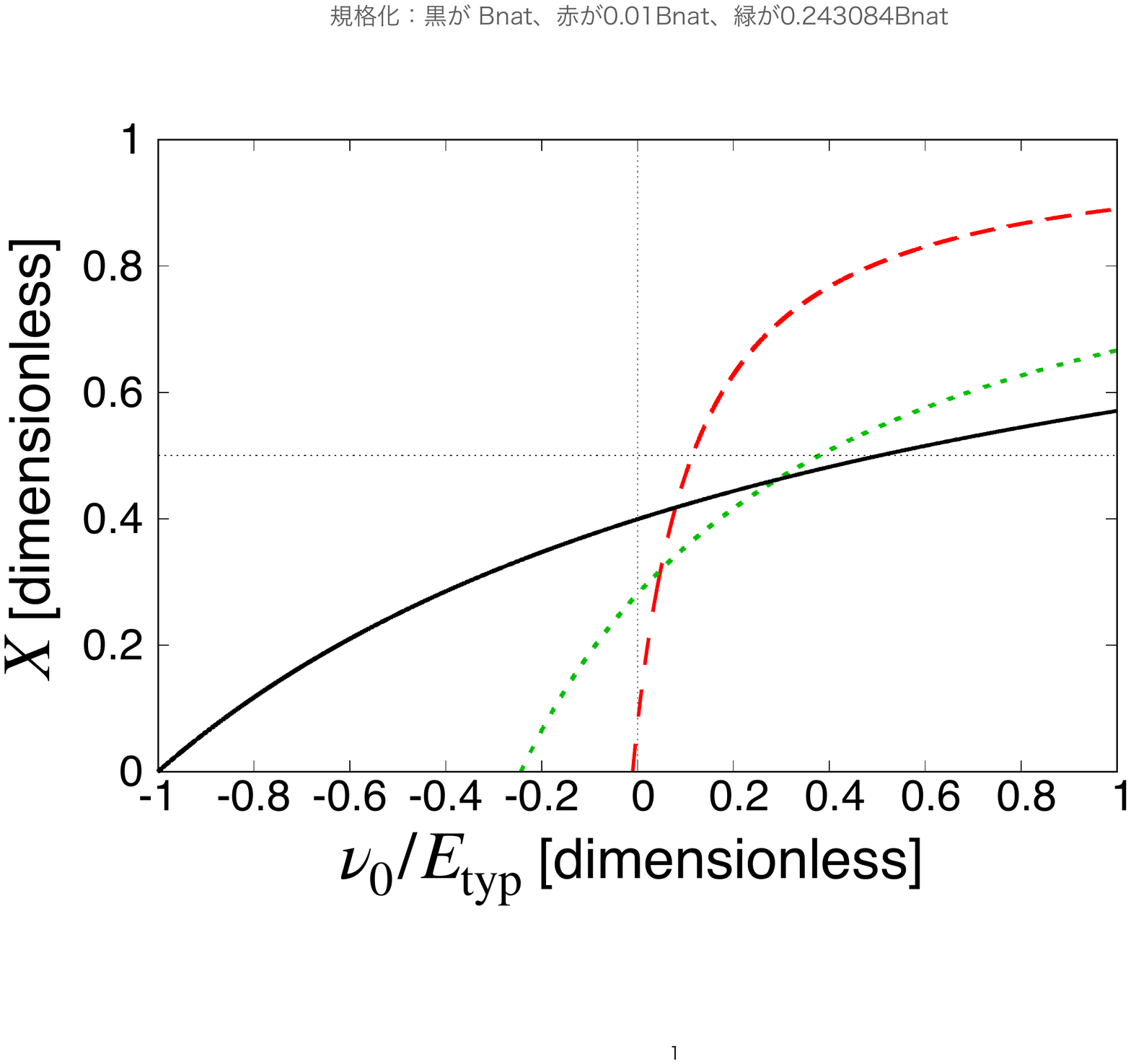}
\caption{The compositeness $X$ as a function of the normalized bare state energy $\nu_{0}/E_{\rm typ}$ with the binding energies $B=E_{\rm typ}$ (solid line), $B=0.01E_{\rm typ}$ (dashed line) and $B=B_{\rm cr}=0.243E_{\rm typ}$ (dotted line). }
\label{fig:X-B}
\end{figure}

In Fig.~\ref{fig:X-B}, we plot the compositeness $X$ as a function of normalized bare state energy $\nu_{0}/E_{\rm typ}$.\footnote{Note that the compositeness $X$ in Eq.~\eqref{eq:X-1ch} depends implicitly on $\nu_{0}$ through $g_{0}^{2}$ [see Eq.~\eqref{eq:g02}].} First, we focus on the solid line which represents $X$ of a bound state with the typical binding energy $B=E_{\rm typ}$. For the most of the allowed region $-1\leq \nu_{0}/E_{\rm typ}\leq 1$, the compositeness $X$ is smaller than 0.5. In other words, the bound state with $B=E_{\rm typ}$ is elementary dominant for most of the $\nu_{0}/E_{\rm typ}$ region. Because the $\nu_{0}$ dependence of the compositeness can be regarded as the model dependence, it is probable one can obtain the bound state with $X<0.5$ in a randomly chosen model. It is consistent with a naive expectation for the model in Sec.~\ref{subsec:eft-2} because the origin of the bound state is the bare state $\phi$ which contributes to the elementarity. 

We then discuss $X$ of a weakly bound state. The dashed line in Fig.~\ref{fig:X-B} corresponds to the case with $B=0.01E_{\rm typ}$ as a representative value of a small binding energy. In this case, the allowed region of $\nu_{0}/E_{\rm typ}$ in Eq.~\eqref{eq:nu0-region} is $-0.01\leq \nu_{0}/E_{\rm typ}\leq 1$. In contrast to the typical bound state with $B=E_{\rm typ}$, $X$ is larger than 0.5 for most of the allowed region of $\nu_{0}/E_{\rm typ}$. Therefore, with the assumption of naturalness, the weak-binding state is mostly composite dominant, even though the bound state originates from the bare state.\footnote{As discussed around Eq.~\eqref{eq:Etyp}, the naturalness requires $B\sim E_{\rm typ}$. To obtain the weakly bound state with $B\ll E_{\rm typ}$, one needs to fine tune of the parameters of the model~\cite{vanKolck:2022lqz}. Here we consider naturalness in the presence of the weakly bound state; namely, we assume that there is no further fine tuning ($\nu_{0}\sim -B$) on top of $B\ll E_{\rm typ}$.} A similar observation was made in Ref.~\cite{Lebed:2022vks} where the internal structure of exotic hadrons was discussed in the dynamical diquark model through the coupling of the tetraquark to the two-meson states. They show that the near-threshold states are dominated by the meson molecular component while the states far from the threshold show tetraquark dominance. When we focus on the $\nu_{0}\sim -B$ region, however, the compositeness of the shallow bound state is small. This is because the compositeness is fixed to be zero in the $\nu_{0}\to -B$ limit. This means that we can always generate an elementary dominant state by choosing the bare state energy appropriately. However, we need a fine tuning of $\nu_{0}$ in the small region around $\nu_{0}\sim -B$ to realize an elementary dominant state. From naturalness, such a fine tuning is unlikely. This is also shown with the separable potential model and the model which is designed to show the scattering effects of a confined state in Ref.~\cite{Hanhart:2014ssa}. In summary, for shallow bound states, the probability of realizing the composite dominant state is much higher than the elementary dominant case, although the latter possibility cannot be completely excluded.

At $B=0$, because the coupling constant in Eq.~\eqref{eq:g02} becomes finite [$g_0^2=\pi^2 \nu_0/(\mu \Lambda)$], from Eq.~\eqref{eq:X-1ch}, $X=1$ holds in the whole region of $0\leq \nu_{0}/E_{\rm typ}\leq 1$. Therefore, the plot of the compositeness in Fig.~\ref{fig:X-B} becomes a step function. This is understood from the low-energy universality~\cite{Braaten:2004rn, Naidon:2016dpf}. It is known that the compositeness $X$ becomes unity in the weak-binding limit $B\to 0$ (compositeness theorem)~\cite{Hyodo:2014bda}. The present model indeed follows this model independent result. From the expectation of the low-energy universality, the microscopic details such as the value of $\nu_{0}$ become irrelevant, and the same relation $X=1$ holds for all models in the $B\to 0$ limit. In contrast to the finite $B\neq 0$ case, the elementary dominant state cannot be generated with any $\nu_{0}$. By gradually decreasing the binding energy from $B=0.01E_{\rm typ}$ (dashed line in Fig.~\ref{fig:X-B}), the region of $\nu_{0}$ with the elementary dominant state becomes smaller and finally vanishes. 

We search for the critical binding energy $B_{\rm cr}$ at which the fractions of the composite dominant region and the elementary dominant region of $\nu_{0}/E_{\rm typ}$ are precisely half and half. From the numerical calculation, it turns out that $B_{\rm cr}= 0.243E_{\rm typ}$, and we plot the compositeness with $B_{\rm cr}$ as the dotted line in Fig.~\ref{fig:X-B}. Namely, we expect that the composite dominant nature of the bound state becomes prominent for the state with $B<B_{\rm cr}$. However we note that $B_{\rm cr}=0.243E_{\rm typ}$ is a value specific to the present model. The value depends on the choice of the regularization of the function $G$ and the interaction Lagrangian.

As a common feature of all cases shown in Fig.~\ref{fig:X-B}, $X$ increases with $\nu_{0}/E_{\rm typ}$. We can analytically show this behavior from Eq.~\eqref{eq:X-1ch} because $g_{0}^{2}$ monotonically increases with $\nu_{0}$. In the $\nu_{0}\to -B$ limit, the compositeness $X$ vanishes.\footnote{For $\nu_{0}<-B$, the compositeness $X$ becomes negative. This is because the norm of the bare state becomes negative for $g_{0}^{2}<0$ and the admixture of the negative norm bare state gives $Z<0$. However, here we do not consider such cases with non-Hermitian Hamiltonian as discussed above.}
When $g_{0}^{2}$ increases, the bound state couples more strongly to the scattering states, and the contribution of the scattering states, and hence the compositeness $X$, becomes larger. This can also be  seen in the $\nu_{0}$ dependence of $X$ in Eq.~\eqref{eq:X-1ch} mentioned above. This relation between the compositeness and the coupling constant has been discussed in the literature~\cite{Sazdjian:2022kaf,Hanhart:2022qxq}. In Ref.~\cite{Sazdjian:2022kaf}, in a similar model setup, the elementarity $Z$ of a bound state is evaluated, and is shown to decrease as a function of the coupling constant. This is essentially equivalent to the result shown in Fig.~\ref{fig:X-B}. In Ref.~\cite{Hanhart:2022qxq}, the authors discuss the nature of the bound state in the weak- and strong-coupling limits, and conclude that the bound state is elementary dominant (composite dominant) for the weak-coupling (strong-coupling) case. Their result is consistent with our analysis of $\nu_{0}\sim -B$ and $\nu_{0}\sim E_{\rm typ}$ in Fig.~\ref{fig:X-B}. Because $\nu_{0}\sim -B$ ($\nu_{0}\sim E_{\rm typ}$) corresponds to $g_{0}\sim 0$ (large $g_{0}$) as seen in Eq.~\eqref{eq:g02}, states are elementary dominant with $X\sim 0$ (composite dominant with $X\sim 1$) in the weak-coupling (strong-coupling) case in our model.

\begin{figure}
\centering
\includegraphics[width=0.5\textwidth]{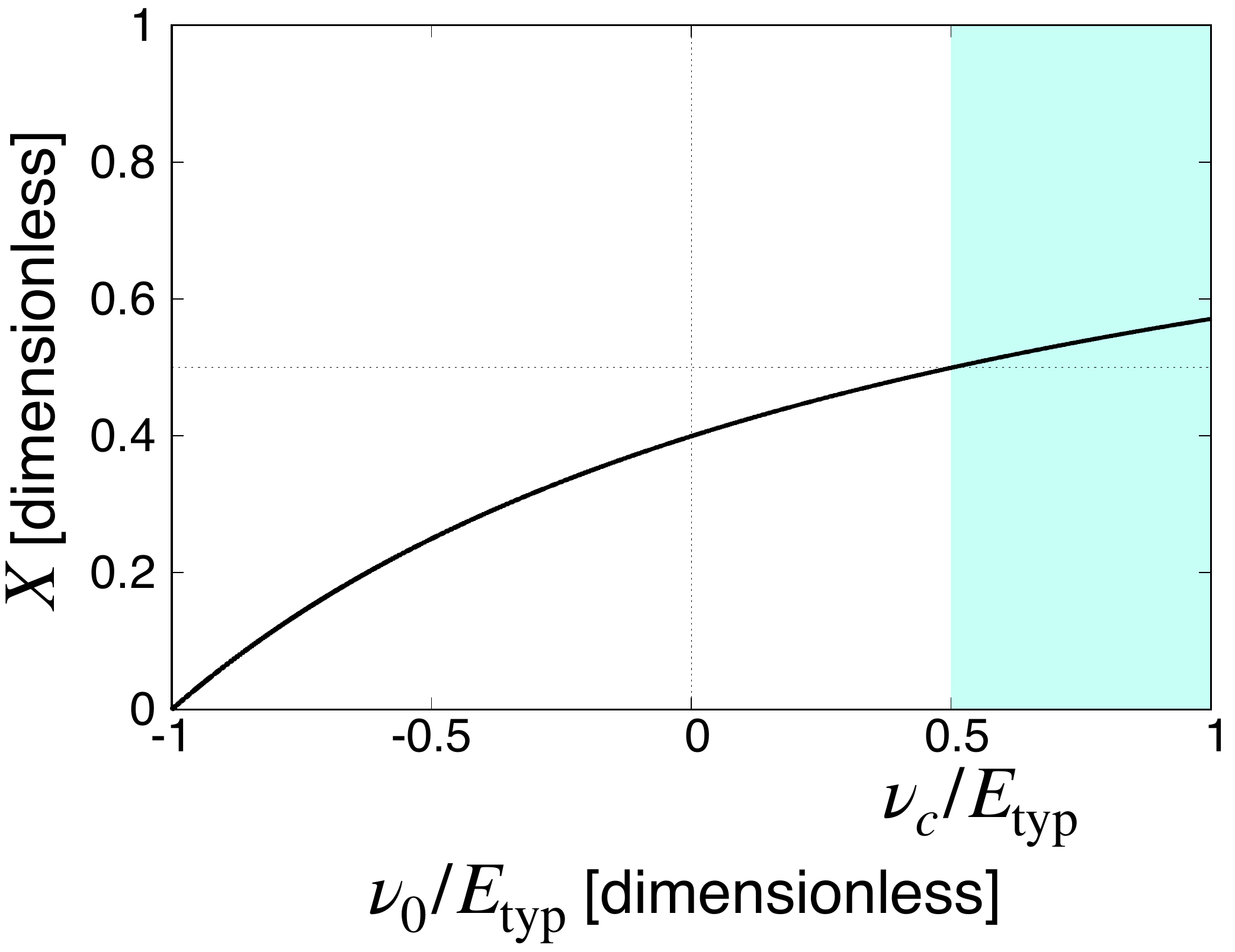}
\caption{The illustration of the definitions of $\nu_{c}$ and $P_{\rm comp}$ with the bound state with $B=E_{\rm typ}$. $P_{\rm comp}$ is the fraction of the shaded region to the entire horizontal axis.}
\label{fig:Pcomp-sch}
\end{figure}

To quantitatively discuss the probability of finding a model with the composite dominant bound state, we define $P_{\rm comp}$ as the fraction of the $\nu_{0}/E_{\rm typ}$ region with $X>0.5$:
\begin{align}
P_{\rm comp}
&=\frac{1-\nu_{c}/E_{\rm typ}}{1+B/E_{\rm typ}},
\label{eq:Pcomp}
\end{align}
where $\nu_{c}$ is the value of $\nu_{0}$ such that $X=0.5$.\footnote{If $X<0.5$ in the whole $\nu_{0}$ region, we set $\nu_{c}=E_{\rm typ}$ and hence $P_{\rm comp}=0 $. In the $B\to 0$ limit, the compositeness is always unity and hence we define $\nu_{c}=-B$ such that $P_{\rm comp}=1 $.} The definition of $\nu_{c}$ for $B=E_{\rm typ}$ is illustrated in Fig.~\ref{fig:Pcomp-sch}. Because $X$ is plotted in the region $-B/E_{\rm typ}\leq \nu_{0}/E_{\rm typ} \leq 1$, the denominator of Eq.~\eqref{eq:Pcomp} corresponds to the length of the horizontal axis in Fig.~\ref{fig:Pcomp-sch}. The numerator is expressed by the width of the shaded region. Thus, $P_{\rm comp}$ in Eq.~\eqref{eq:Pcomp} is defined as the ratio of the shaded region to all of the allowed region of $\nu_{0}/E_{\rm typ}$ (the length of the horizontal axis). The explicit values for the cases shown in Fig.~\ref{fig:X-B} are found to be $P_{\rm comp}=0.25$ at $B=E_{\rm typ}$ and $P_{\rm comp}=0.88$ at $B=0.01E_{\rm typ}$. Because $B_{\rm cr}$ is defined so that the composite dominant case occupies the half of the whole $\nu_{0}/E_{\rm typ}$ region, we obtain $P_{\rm comp}=0.5$ at $B=B_{\rm cr}$.

\begin{figure}
\centering
\includegraphics[width=0.5\textwidth]{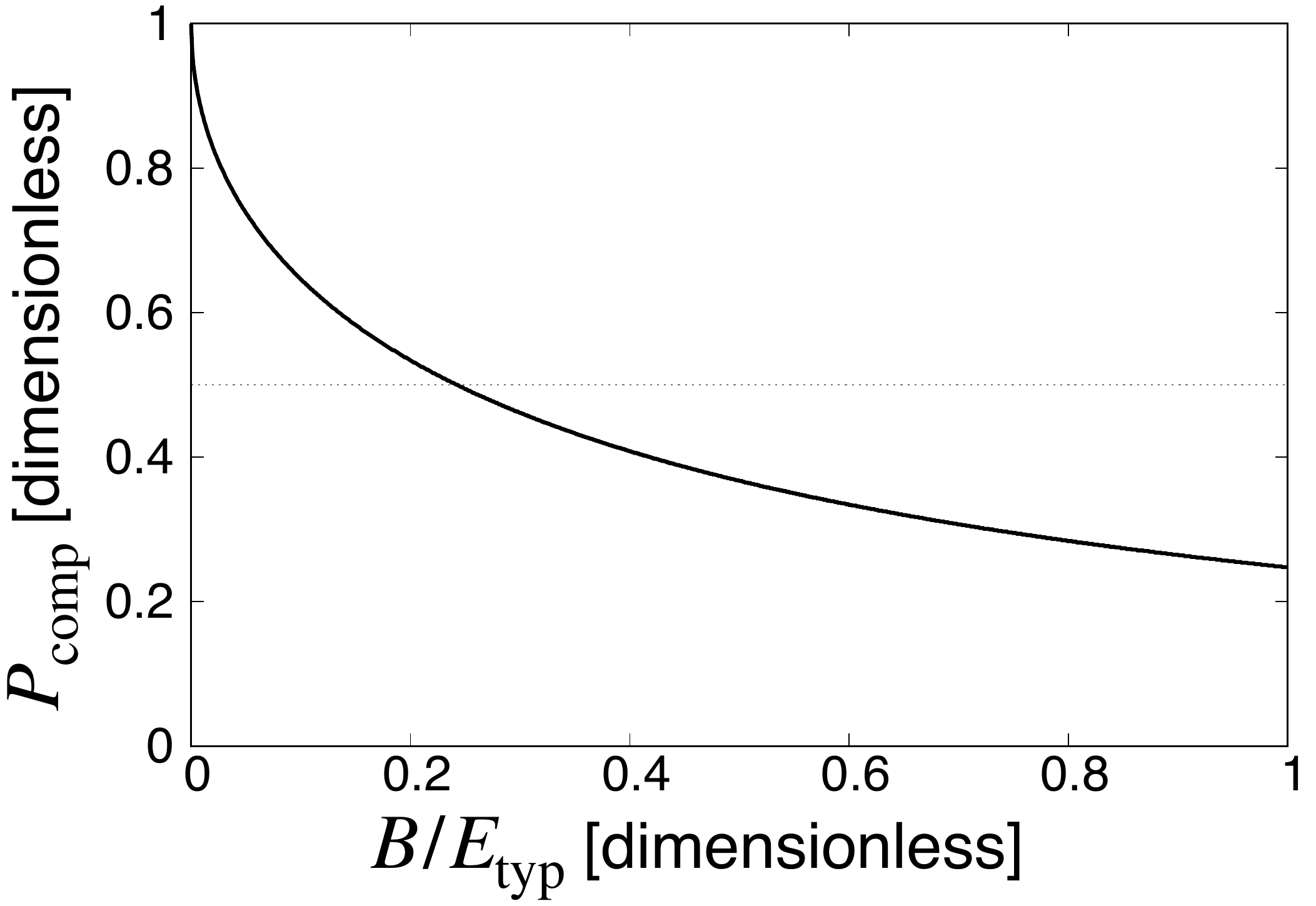}
\caption{The fraction of the composite dominant region $P_{\rm comp}$ as a function of normalized binding energy $B/E_{\rm typ}$. }
\label{fig:B-Pcomp}
\end{figure}

In Fig.~\ref{fig:B-Pcomp}, we plot $P_{\rm comp}$ by varying the normalized binding energy $B/E_{\rm typ}$. Small $P_{\rm comp}$ at $B/E_{\rm typ}\sim 1$ monotonically increases to unity by decreasing the binding energy $B$. This shows that the probability of finding a model with the composite dominant state is small for the typical bound states but gradually increases along with the reduction of the binding energy. In the small $B$ region, $P_{\rm comp}$ rapidly grows toward unity. At $B=0$, we have $P_{\rm comp}=1$; the bound state becomes completely composite dominant for any model, as discussed above. Figure~\ref{fig:B-Pcomp} also shows that, even when we slightly go away from the weak-binding limit $B=0$, it is expected that $P_{\rm comp}$ is still close to unity. This suggests that it is probable to find the near-threshold composite dominant states with $B\ll E_{\rm typ}$. In fact, the weak-binding hadrons, nuclei, and atoms studied in Ref.~\cite{Kinugawa:2022fzn} are all composite dominant states.

\subsection{Weak-binding relation}
\label{subsec:wbr}
Finally, we discuss the results in the previous section by comparing with the compositeness in the zero-range limit in Eq.~\eqref{eq:wbs}. For this purpose, let us consider the idealized case when the interaction range is sufficiently small. As shown in Refs.~\cite{Weinberg:1965zz,Kamiya:2016oao,Oller:2017alp,Kinugawa:2022fzn}, the weak-binding relation gives the compositeness $X_{\rm wb}$ from the scattering length $a_{0}$ and the radius of the bound state $R=1/\kappa$ as follows:
\begin{align}
X_{\rm wb}&=X_{\rm wb}^{c}+\mathcal{O}\left(\frac{R_{\rm typ}}{R}\right), 
\label{eq:wbr}\\
X_{\rm wb}^{c}&=\frac{a_{0}}{2R-a_{0}},
\quad
R_{\rm typ}=\max\{1/\Lambda, |r_{e}|,...\}, \label{eq:wbrc}
\end{align}
where $r_{e}$ is the effective range, $R_{\rm typ}$ is the typical length scale of the system which is estimated as the maximum length scale expected for $a_{0}$, and $X_{\rm wb}^{c}$ is the central value of the compositeness in the weak-binding relation. For a weakly bound state with $R\gg R_{\rm typ}$, we can neglect the correction terms $\mathcal{O}(R_{\rm typ}/R)$ in Eq.~\eqref{eq:wbr}. In this case, $X_{\rm wb}\sim X_{\rm wb}^{c}$ is obtained only from the observables $a_{0}$ and $R$. This is equivalent to Eq.~\eqref{eq:wbs} in the zero-range limit. Therefore, the weak-binding relation is a model independent method to estimate the compositeness of the shallow bound state. 

\begin{figure*}
\centering
\includegraphics[width=0.45\textwidth]{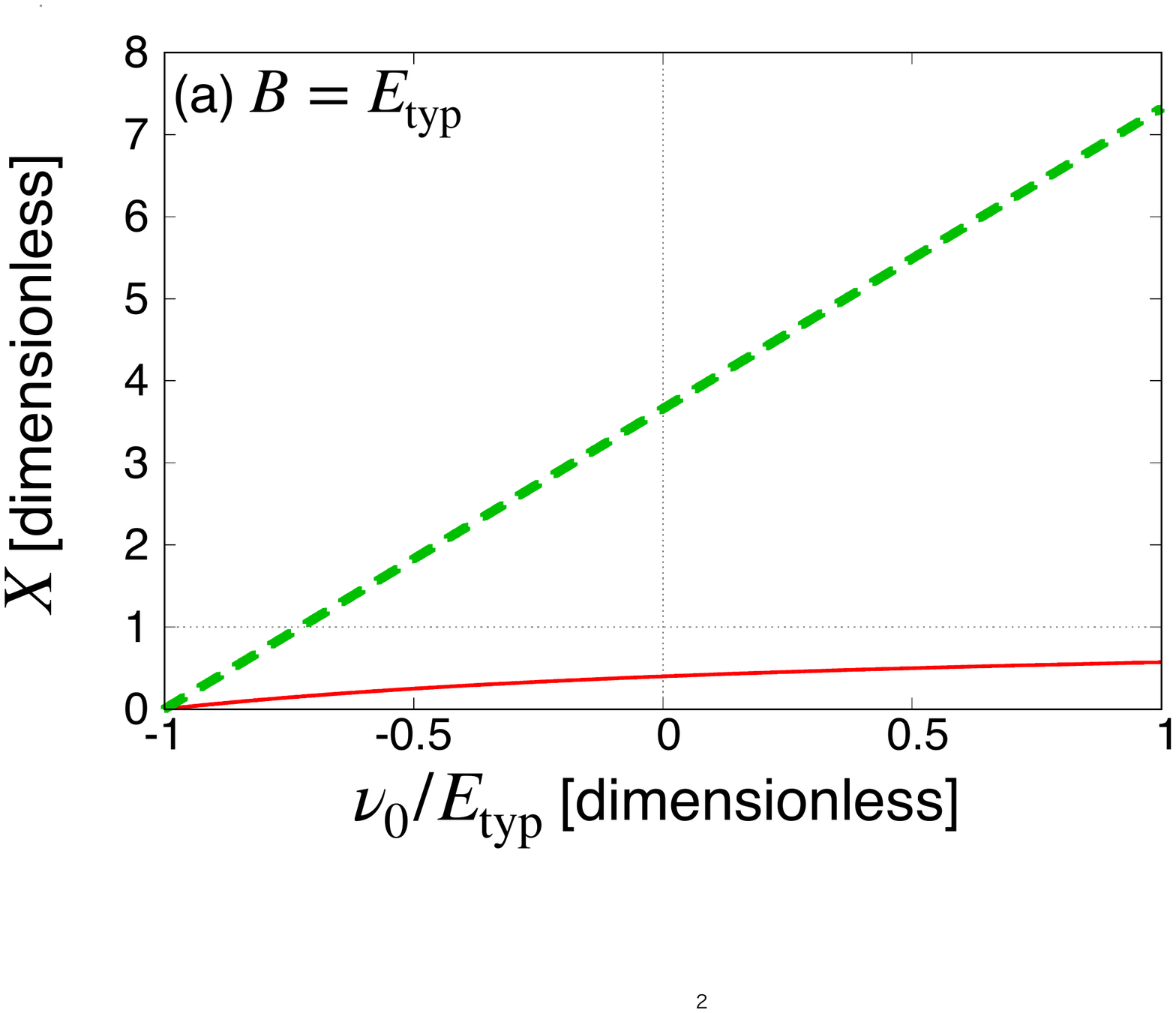}
\includegraphics[width=0.45\textwidth]{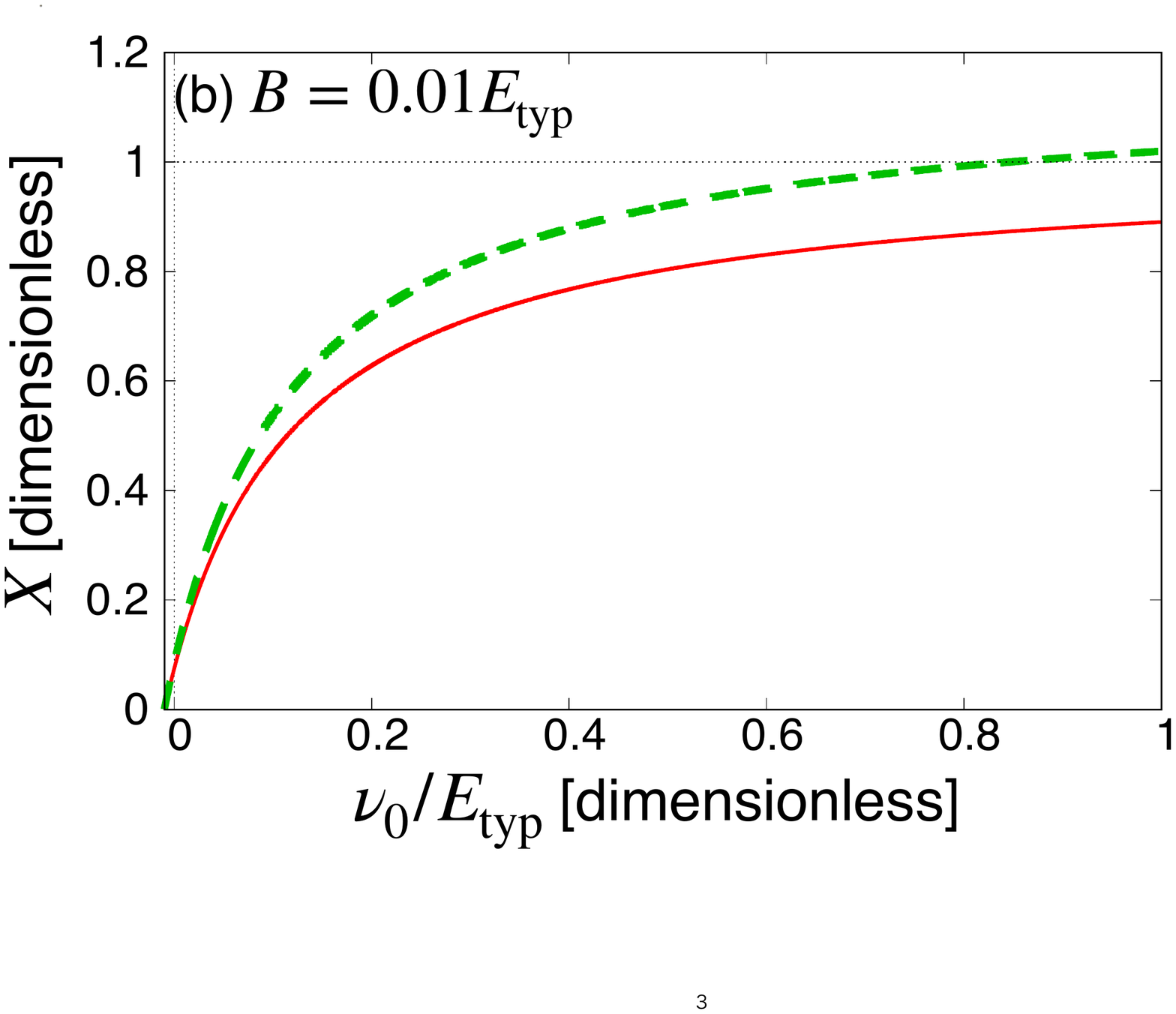}
\caption{The compositeness obtained from the model calculation~\eqref{eq:X-1ch} (solid line) and from the central value of the weak-binding relation~\eqref{eq:wbrc} (dashed line) with the fixed binding energy $B=E_{\rm typ}$ [panel (a)] and $B=0.01E_{\rm typ}$ [panel (b)].}
\label{fig:wbr}
\end{figure*}

In Fig.~\ref{fig:wbr}, we plot the compositeness $X$ in this model [Eq.~\eqref{eq:X-1ch}] and the central value of the compositeness from the weak-binding relation $X_{\rm wb}^{c}$ in Eq.~\eqref{eq:wbrc}. We use $a_{0}$ in Eq.~\eqref{eq:1ch-a} and $R$ in this model for the calculation of $X_{\rm wb}^{c}$. Panel (a) shows $X$ (solid line) and $X_{\rm wb}^{c}$ (dashed line) for the typical binding case $B=E_{\rm typ}$, and panel (b) similarly for the weak-binding case $B=0.01E_{\rm typ}$. We see that the difference between $X_{\rm wb}^{c}$ and $X$ is significant for $B=E_{\rm typ}$ in panel (a) while that for $B=0.01E_{\rm typ}$ is at most 0.1 in panel (b). Therefore, the weak-binding relation gives a good estimation of the compositeness for the shallow bound state. It is remarkable that the weak-binding relation works to estimate $X$ correctly even in the region $\nu_{0}\sim -B$ where $X<0.5$ in panel (b). In Ref.~\cite{Kinugawa:2022fzn}, the validity of the weak-binding relation is demonstrated for composite dominant ($X\sim 1$) states with a shallow binding energy. In this work, we find that the weak-binding relation works also for shallow but elementary dominant bound states.

It is instructive to analytically show that the exact compositeness $X$ coincides with the weak-binding one in Eq.~\eqref{eq:wbrc} in the small $B$ limit. For a weakly bound state with $B\ll E_{\rm typ}$ ($\kappa\ll \Lambda$), the arctangent term in the loop function in Eq.~\eqref{eq:G-1ch} can be approximated as
\begin{align}
\arctan\left(\frac{\Lambda}{\kappa}\right)&=\frac{\pi}{2}+\mathcal{O}\left(\frac{\kappa}{\Lambda}\right).
\label{eq:approximation}
\end{align}
Under this approximation, the loop function $G(i\kappa)$ becomes
\begin{align}
G(i\kappa)\approx-\frac{\mu}{\pi^{2}}\left(\Lambda-\frac{\pi}{2}\kappa\right)
\quad (B\ll E_{\rm typ}) ,
\label{eq:approx-G}
\end{align}
and the scattering amplitude is given by
\begin{align}
f(i\kappa)&\approx\left[\frac{2\pi}{\mu}\left(\frac{\frac{\kappa^{2}}{2\mu}+\nu_{0}}{g_{0}^{2}}\right)-\frac{2\Lambda}{\pi}+\kappa\right]^{-1}
\quad (B\ll E_{\rm typ})
.
\label{eq:amplitude-approx}
\end{align}
Note that the approximation in Eq.~\eqref{eq:approximation} is not valid for a large binding energy ($\kappa >2\Lambda/\pi$) where the square of the coupling constant becomes negative, $g^{2}_{0}<0$.
From $V$ and the approximated $G$ in Eq.~\eqref{eq:approx-G}, the compositeness $X$ in Eq.~\eqref{eq:X-1ch} is given by
\begin{align}
X\approx\left[1+\frac{2\pi}{R\mu^{2}g_{0}^{2}}\right]^{-1}
\quad (B\ll E_{\rm typ}).
\label{eq:X-approx}
\end{align}
By using the scattering length $a_{0}$ in Eq.~\eqref{eq:1ch-a}, the central value of the compositeness in Eq.~\eqref{eq:wbrc} is obtained as
\begin{align}
X_{\rm wb}^{c}&=\left[2R\left(-\frac{2\pi\nu_{0}}{g_{0}^{2}\mu}+\frac{2}{\pi}\Lambda\right)-1\right]^{-1}\nonumber \\
&=\left[2R\left(\frac{\pi}{R^{2}g_{0}^{2}\mu^{2}}+\frac{1}{R}\right)-1\right]^{-1}\nonumber \\
&=\left[1+\frac{2\pi}{R\mu^{2}g_{0}^{2}}\right]^{-1}.
\label{eq:Xwb-approx}
\end{align}
In the second line, we use the bound state condition from the scattering amplitude in Eq.~\eqref{eq:amplitude-approx} written using $R=1/\kappa$:
\begin{align}
-\frac{2\pi}{\mu}\left(\frac{-\frac{\kappa^{2}}{2\mu}-\nu_{0}}{g_{0}^{2}}\right)-\frac{2\Lambda}{\pi}+\kappa&=0\nonumber \\
\Leftrightarrow \frac{2\Lambda}{\pi}-\frac{2\pi\nu_{0}}{g_{0}^{2}\mu}=\frac{\pi}{R^{2}g_{0}^{2}\mu^{2}}+\frac{1}{R}.
\end{align}
From Eqs.~\eqref{eq:X-approx} and \eqref{eq:Xwb-approx}, we show that the exact compositeness $X$ reduces to the central value estimated by the weak-binding relation $X_{\rm wb}^{c}$ in the small $B$ limit. In Ref.~\cite{Kamiya:2016oao}, it is shown that there are two origins of the deviation of estimated $X_{\rm wb}^{c}$ from exact $X$. The first one comes from the higher order terms in the derivative of the loop function, and the second one from those in the effective range expansion of the residue of the bound state pole. The derivative of the approximated loop function in Eq.~\eqref{eq:approx-G} has only the leading order term, and hence the first deviation does not appear. Because the scattering amplitude in Eq.~\eqref{eq:amplitude-approx} has no higher order terms of $\mathcal{O}(k^{4})$, the second deviation does not arise. In this way, we explicitly show that all the deviations disappear in the $B\to 0$ limit and the estimation of the compositeness using the weak-binding relation becomes exact. In this context, it is worth noting the deviation of $X_{\rm wb}^c$ from exact $X$ in the scattering models discussed in Ref.~\cite{Braaten:2007nq,Kinugawa:2022fzn}. In the zero-range model with the loop function in Eq.~\eqref{eq:approx-G}, $X=X_{\rm wb}^{c}$ can be shown because the inverse scattering amplitude is given up to $\mathcal{O}(k)$. In contrast, $X$ deviates from $X_{\rm wb}^{c}$ in the resonance model because the four-point contact interaction induces the higher order terms of $\mathcal{O}(k^{4})$ in the effective range expansion. 


\section{Effects of four-point contact interaction, decay, and channel coupling}
\label{sec:effects}

As mentioned in the Introduction, actual exotic hadrons have finite decay width and coupling to the additional scattering channel. One can also consider the direct interaction in the threshold channel which is absent in the model in Sec.~\ref{sec:wbs}. In this section, we consider the four-point contact interaction, decay, and coupled-channels effect and show how these contributions modify the results in the previous section. In Sec.~\ref{subsec:contact}, we introduce the four-point contact interaction in addition to the model in the previous section, and study the contribution of the four-point interaction to the compositeness and low-energy universality. In the same way, the decay contribution and coupled-channels contribution are discussed in Secs.~\ref{subsec:decay} and \ref{subsec:2ch}, respectively. 

\subsection{Effect of four-point contact interaction}
\label{subsec:contact}

In this section, we investigate the effect of the direct interaction of $\psi_{1}$ and $\psi_{2}$ in addition to the model in Sec.~\ref{sec:wbs}. For this purpose, we introduce the four-point contact interaction term with the coupling constant $\lambda_{0}$, and the interaction Hamiltonian in Eq.~\eqref{eq:H-wbs} becomes
\begin{align}
 \mathcal{H}_{\rm int}&=\lambda_{0}(\psi_{1}^{\dagger}\psi_{2}^{\dagger}\psi_{1}\psi_{2})+g_{0}(\phi^{\dagger}\psi_{1}\psi_{2}+\psi^{\dagger}_{1}\psi^{\dagger}_{2}\phi).
 \label{eq:contact-H}
\end{align}
Positive $\lambda_{0}>0$ (negative $\lambda_{0}<0$) corresponds to a repulsive (attractive) interaction. Because of the addition of the contact interaction term, the effective interaction $V(k)$ in Eq.~\eqref{eq:V-1ch} as a function of the momentum $k$ changes to 
\begin{align}
V(k)&=\lambda_{0}+\frac{g_{0}^{2}}{\frac{k^{2}}{2\mu}-\nu_{0}},
\label{eq:V-lambda}
\end{align}
while the loop function $G(k)$ in Eq.~\eqref{eq:G-1ch} remains unchanged. The scattering amplitude $f(k)$ is obtained as
\begin{align}
f(k)&=-\frac{\mu}{2\pi}\left[\left(\lambda_{0}+\frac{g_{0}^{2}}{\frac{k^{2}}{2\mu}-\nu_{0}}\right)^{-1} \right.\nonumber \\
& \quad \left.+\frac{\mu}{\pi^{2}}\left\{\Lambda+ik\arctan\left(-\frac{\Lambda}{ik}\right)\right\}\right]^{-1}. 
\label{eq:contact-amplitude}
\end{align}
As in Sec.~\ref{subsec:eft-2}, we consider the bound state with the eigenmomentum $k=i\kappa$ and the binding energy $B=\kappa^{2}/(2\mu)$. The compositeness $X$ is calculated from $V$ in Eq.~\eqref{eq:V-lambda} and $G$ in Eq.~\eqref{eq:G-int}:
\begin{align}
X&=\left[1+\frac{\frac{g_{0}^{2}\pi^{2}\kappa}{\mu^{2}}}{{(B+\nu_{0})^{2}\left(\lambda_{0}-\frac{g_{0}^{2}}{B+\nu_{0}}\right)^{2}}\left(\arctan\left(\frac{\Lambda}{\kappa}\right)-\frac{\frac{\Lambda}{\kappa}}{1+\left(\frac{\Lambda}{\kappa}\right)^{2}}\right)}\right]^{-1}.
\label{eq:X-lambda}
\end{align}

The model parameters are the bare state energy $\nu_{0}$, the cutoff $\Lambda$, and the coupling constants $\lambda_{0}$ and $g_{0}$. As in the model in Sec.~\ref{subsec:eft-2}, from the bound state condition $f(i\kappa)^{-1}=0$ with a fixed binding energy, $g_{0}^{2}$ is written in terms of the binding energy $B$ and other model parameters:
\begin{align}
g_{0}^{2}(B; \nu_{0},\lambda_{0},\Lambda)&=(B+\nu_{0})\left(\frac{\frac{\pi^{2}}{\mu}}{\Lambda-\kappa\arctan\left(\frac{\Lambda}{\kappa}\right)}+\lambda_{0}\right).
\label{eq:g02-lambda0}
\end{align}
Furthermore, the use of the dimensionless parameters can absorb the $\Lambda$ dependence. Therefore, the remaining parameters $\nu_{0}$ and $\lambda_{0}$ are varied in the calculation of the compositeness. Namely, $\nu_{0}$ and $\lambda_{0}$ dependence of the compositeness can be regarded as the model dependence in this case. As in Sec.~\ref{subsec:calc-2}, we vary $\nu_{0}/E_{\rm typ}$ in the region $-B/E_{\rm typ}\leq \nu_{0}/E_{\rm typ} \leq 1$.\footnote{As in Sec.~\ref{sec:wbs}, the lower boundary of $\nu_{0}$ is determined by the condition $g_{0}^{2}>0$. Note that even with the choice of $\nu_{0}$ in the $\nu_{0} < -B$ region, $g_{0}^{2}$ can still be positive if $\lambda_{0}$ is sufficiently large and negative. However, we exclude such a case because $|\lambda_{0}|$ is restricted as discussed below. In this study, we vary $|\lambda_{0}|$ up to $|\lambda_{0}^{\rm cr}|$ in Eq.~\eqref{eq:lambda-cr} which is regarded as the typical value of the strong coupling. In this case, the condition $-B\leq \nu_{0}$ should be satisfied to obtain positive $g_{0}^{2}$. }

We now consider the relevant parameter region of $\lambda_{0}$. From Eq.~\eqref{eq:g02-lambda0}, we see that $g_{0}^{2}$ becomes negative for large negative $\lambda_{0}$. To avoid this problem, we introduce the lower boundary of $\lambda_{0}$ as $\lambda_{0}^{\rm b}$, which is determined by the condition $g_{0}^{2}=0$ in Eq.~\eqref{eq:g02-lambda0}:
\begin{align}
\lambda^{\rm b}_{0}=-\frac{\pi^{2}}{\mu}\left[\Lambda-\kappa\arctan\left(\frac{\Lambda}{\kappa}\right)\right]^{-1}. \label{eq:lambda-b}
\end{align}
Thus, $\lambda_{0}$ should be chosen in the allowed region $-|\lambda_{0}^{\rm b}| \leq \lambda_{0}$. Note that $\lambda^{\rm b}_{0}$ depends on the binding energy $B=\kappa^{2}/(2\mu)$.

We then determine the region of $\lambda_{0}$ for the numerical calculation. We define $\lambda_{0}^{\rm cr}$ as the critical value of the attractive coupling constant which supports a bound state at $B=0$ without the bare state contribution:
\begin{align}
\lambda^{\rm cr}_{0}=-\frac{\pi^{2}}{\mu\Lambda}. \label{eq:lambda-cr}
\end{align}
In fact, with $\lambda_{0}=\lambda^{\rm cr}_{0}$ and $g_{0}^{2}=0$, the scattering amplitude in Eq.~\eqref{eq:contact-amplitude} has a pole at $B=\kappa=0$. With a stronger attraction than $\lambda_{0}^{\rm cr}$, a new bound state is generated by the four-point interaction in addition to the one developed from the bare state. This is qualitatively different from the situation in Sec.~\ref{sec:wbs} where only one bound state exists below the threshold.  Therefore, for the attractive interaction, we impose the condition $\lambda_{0}$ in $-|\lambda_{0}^{\rm cr}| \leq \lambda_{0}$. Because the formation of a bound state is a non-perturbative phenomenon, $|\lambda_{0}^{\rm cr}|$ can be regarded as the representative strength of the strong coupling. Based on this consideration, we restrict $\lambda_{0} \leq |\lambda_{0}^{\rm cr}|$ also for the repulsive interaction. To examine the effect of the four-point interaction to the result of Sec.~\ref{sec:wbs}, we vary $\lambda_{0}$ in the region 
\begin{align}
-|\lambda_{0}^{\rm cr}|\leq \lambda_{0}\leq |\lambda_{0}^{\rm cr}|.
\label{eq:lambda0region}
\end{align} 
Note that the relation $-|\lambda_{0}^{\rm b}|<-|\lambda_{0}^{\rm cr}|$ always holds since $\kappa\arctan(\kappa/\Lambda)>0$ for any $\kappa>0$. Thus, the condition $-|\lambda_{0}^{b}|<\lambda_{0}$ is guaranteed with Eq.~\eqref{eq:lambda0region} 
for any $\kappa$.

\begin{figure}
\centering
\includegraphics[width=0.5\textwidth]{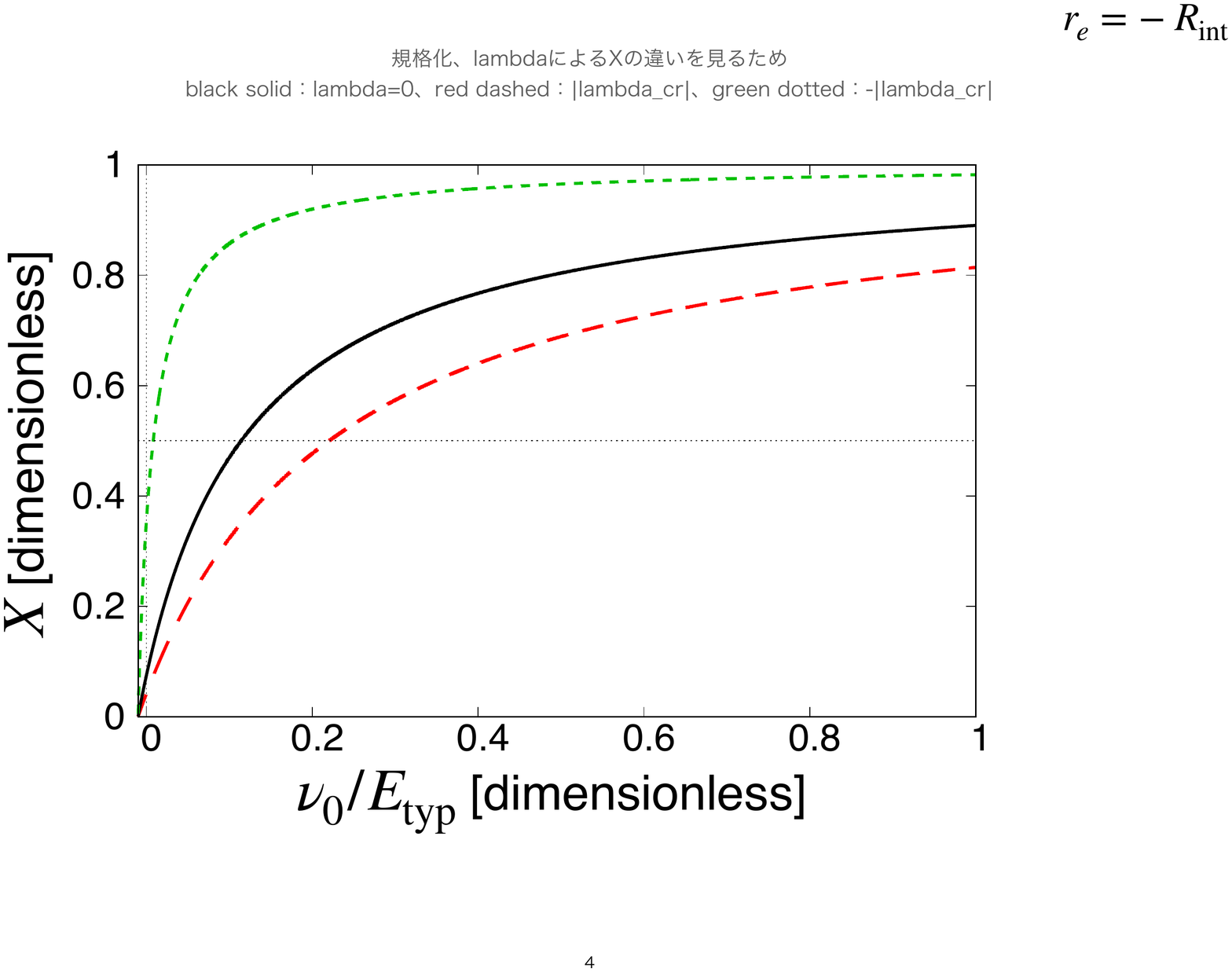}
\caption{The compositeness $X$ as a function of the normalized bare state energy $\nu_{0}/E_{\rm typ}$ for $\lambda_{0}=0$ (solid line), $\lambda_{0}=-|\lambda^{\rm cr}_{0}|$ (dotted line), and $\lambda_{0}=|\lambda^{\rm cr}_{0}|$ (dashed line) with $B=0.01E_{\rm typ}$.}
\label{fig:lambda_0s}
\end{figure}

To observe the effect of the contact interaction with $\lambda_{0}$, we plot the compositeness $X$ as a function of the normalized bare states energy $\nu_{0}/E_{\rm typ}$ for the weak-binding case $B=0.01E_{\rm typ}$ in Fig.~\ref{fig:lambda_0s}. The solid, dashed, and dotted lines express $X$ with $\lambda_{0}=0$, $\lambda_{0}=-|\lambda^{\rm cr}_{0}|$, and $\lambda_{0}=|\lambda^{\rm cr}_{0}|$, respectively. As shown in Fig.~\ref{fig:lambda_0s}, the repulsive interaction $|\lambda^{\rm cr}_{0}|>0$ decreases $X$ and the attractive interaction $-|\lambda^{\rm cr}_{0}|<0$ increases $X$ for fixed $\nu_{0}$. 
To understand this behavior, we consider the interaction mechanisms and their implication for the compositeness of the bound state. In the present model, the bound state originates not only from the bare state pole but also from the attractive four-point interaction.
As discussed in Sec.~\ref{sec:wbs}, the bare pole term contributes to the elementarity $Z$. In contrast, the attractive four-point interaction provides the composite bound state, and hence contributes to the compositeness $X$. With both the interactions, the compositeness of the bound state is determined by the interplay between the bare pole term proportional to $g_{0}^{2}$ and the direct interaction proportional to $\lambda_{0}$. Because the binding energy is chosen to satisfy $-B< \nu_0$, Eq.~\eqref{eq:g02-lambda0} indicates that $g_{0}^{2}$ increases with $\lambda_{0}$ for fixed $B$ and $\nu_{0}$.
Intuitively, negative $\lambda_{0}$ (attractive four-point interaction) tends to increase the binding energy, and hence the coupling to the bare pole term $g_{0}^{2}$ should be reduced to keep the binding energy unchanged. In contrast, $g_{0}^{2}$ increases to compensate for the reduction of the binding energy by the repulsive four-point interaction with positive $\lambda_{0}$. This relation between $g_{0}^{2}$ and $\lambda_{0}$, together with the origin of the bound state discussed above, explains the behavior of the compositeness with respect to $\lambda_{0}$; the introduction of the repulsive (attractive) four-point interaction with positive (negative) $\lambda_{0}$ increases (decreases) $g_{0}^{2}$ and therefore the compositeness of the bound state $X$ decreases (increases).

\begin{figure}
\centering
\includegraphics[width=0.45\textwidth]{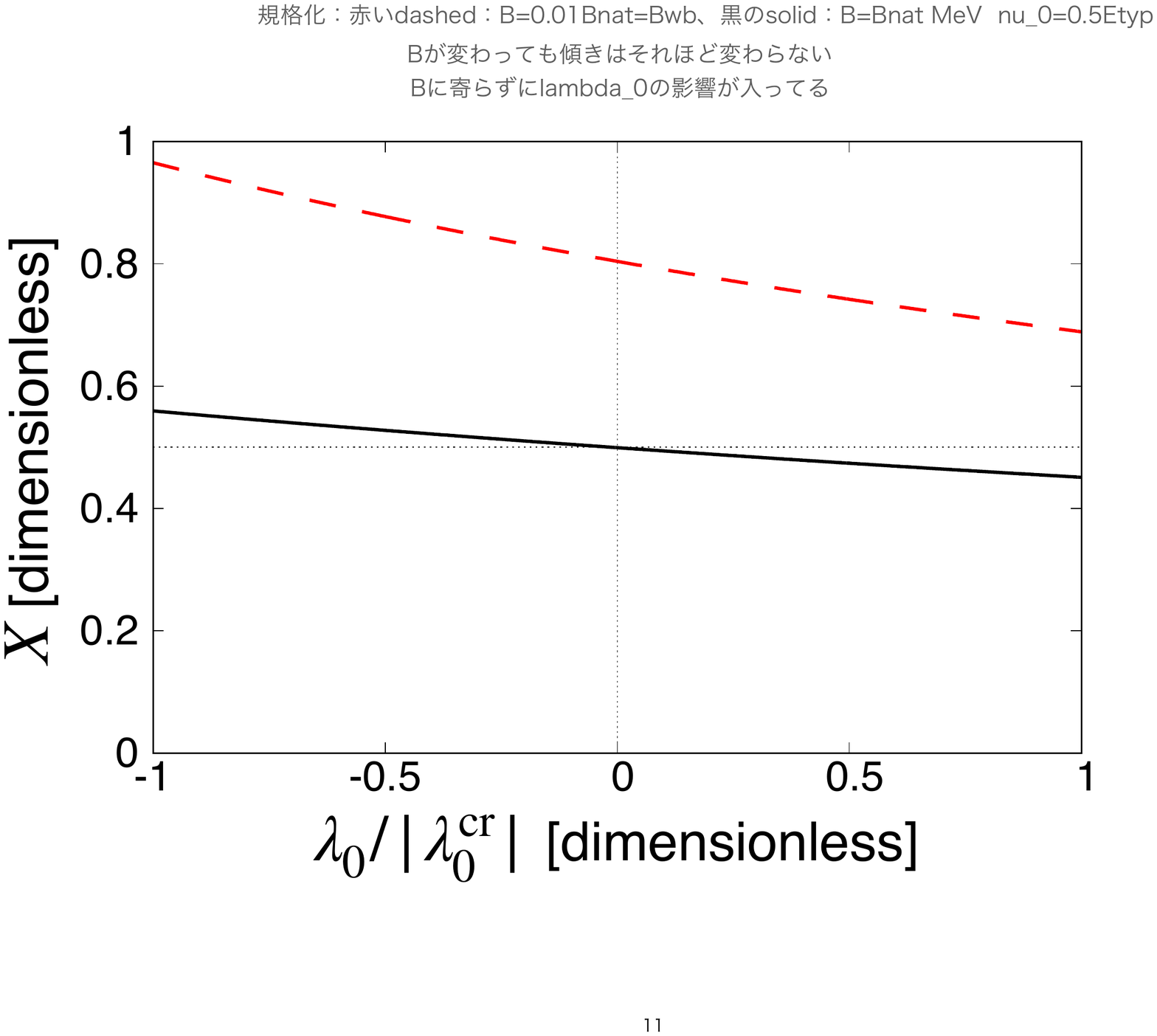}
\caption{The compositeness $X$ as a function of the normalized coupling constant of the four-point interaction $\lambda_{0}/|\lambda_{0}^{\rm cr}|$ for $B=E_{\rm typ}$ (solid line) and $B=0.01 E_{\rm typ}$(dashed line). The bare state energy $\nu_{0}$ is fixed as $\nu_{0}=0.5E_{\rm typ}$.
\label{fig:X-lambda0}}
\end{figure}

The $\lambda_{0}$ dependence of the compositeness can be visualized by plotting $X$ as a function of the normalized  coupling constant $\lambda_{0}/|\lambda_{0}^{\rm cr}|$ in Fig.~\ref{fig:X-lambda0}. In this plot, we fix the bare state energy as $\nu_{0}=0.5E_{\rm typ}$ and we have checked that the qualitative result does not change for different values of $\nu_{0}$. The solid line represents $X$ for $B=E_{\rm typ}$, and the dashed line represents for $B=0.01E_{\rm typ}$. In both cases, $X$ decreases with the increase of $\lambda_{0}/|\lambda_{0}^{\rm cr}|$, as discussed above. In Fig.~\ref{fig:X-lambda0}, we see that the compositeness $X$ depends on $\lambda_{0}$ more strongly for $B=0.01E_{\rm typ}$ than that for $B=E_{\rm typ}$. This tendency originates in the structure of the bound state at $\lambda_{0}=0$; a stronger coupling $g_{0}$ is required to generate the deeper bound state with the same $\nu_{0}$, as indicated by the smaller compositeness $X$ for the typical bound state with $B=E_{\rm typ}$. The deeper bound state is less affected by the introduced four-point interaction
and hence the $\lambda_{0}$ dependence becomes milder.

\begin{figure}
\centering
\includegraphics[width=0.5\textwidth]{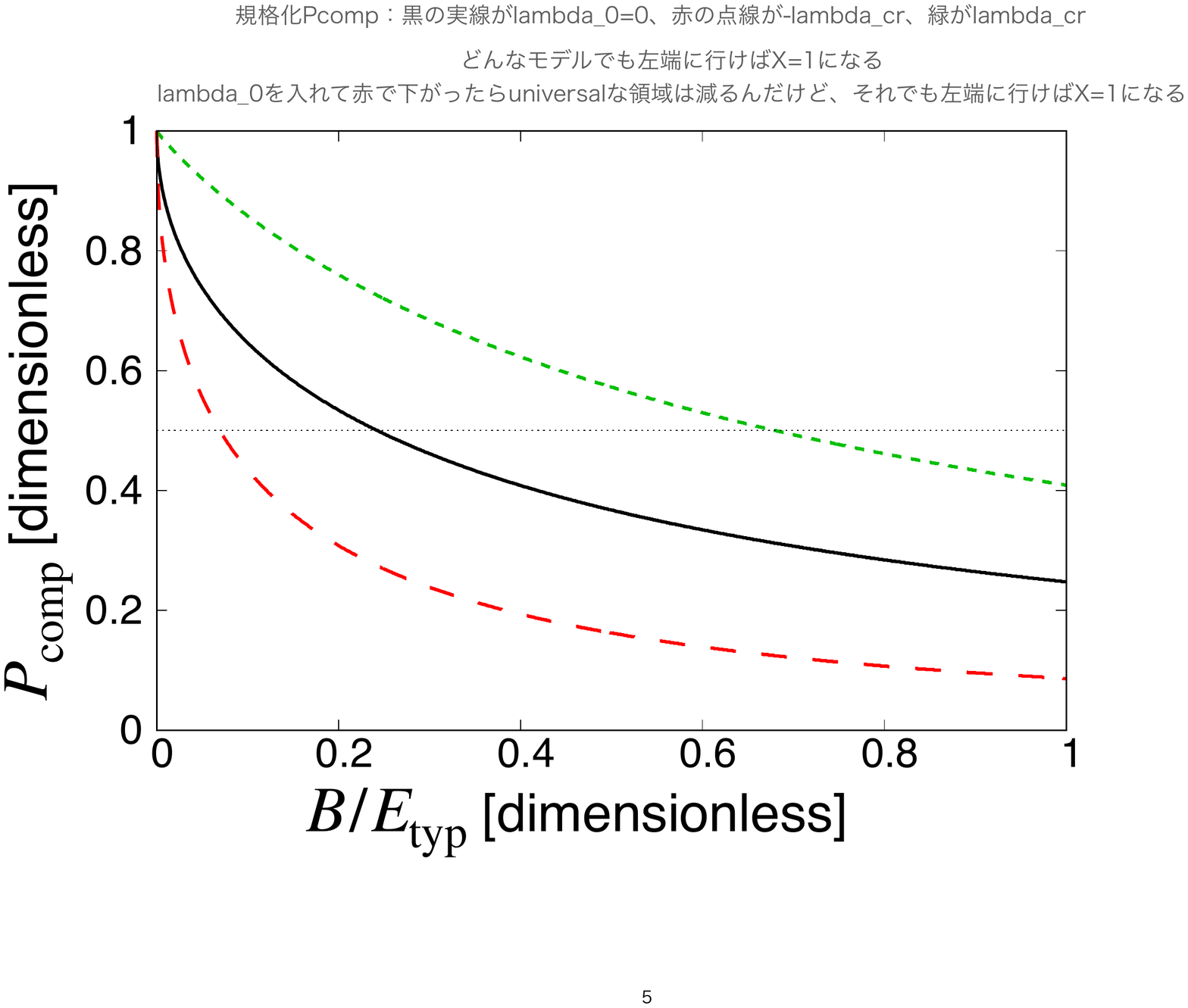}
\caption{The fraction of the composite dominant region $P_{\rm comp}$ as a function of normalized bare state energy $\nu_{0}/E_{\rm typ}$. The solid line stands for $P_{\rm comp}$ for $\lambda_{0}=0$, the dashed line for repulsive $\lambda_{0}=|\lambda_{0}^{\rm cr}|$, and the dotted line for attractive $\lambda_{0}=-|\lambda_{0}^{\rm cr}|$. }
\label{fig:lambda-Pcomp}
\end{figure}

Finally, in Fig.~\ref{fig:lambda-Pcomp}, we plot the fraction of composite dominant state $P_{\rm comp}$ as a function of the normalized binding energy $B/E_{\rm typ}$ to discuss the low-energy universality with $\lambda_{0}$ contribution. The solid line represents $P_{\rm comp}$ with $\lambda_{0}=0$ (same as Fig.~\ref{fig:B-Pcomp}), the dashed line with $\lambda_{0}=|\lambda_{0}^{\rm cr}|$, and the dotted line with $\lambda_{0}=-|\lambda_{0}^{\rm cr}|$. For all the $\lambda_{0}$ cases, $P_{\rm comp}$ decreases when the binding energy $B$ increases. Because positive $\lambda_{0}>0$ (repulsive interaction) suppresses the compositeness and $\nu_{c}$ becomes smaller (see Fig.~\ref{fig:lambda_0s}), $P_{\rm comp}$ is also suppressed. In contrast, attractive interaction with negative $\lambda_{0}<0$ enhances $P_{\rm comp}$ because it induces the increase of $X$ and the decrease of $\nu_{c}$ in Fig.~\ref{fig:lambda_0s}. At $B=0$, we see that $P_{\rm comp}$ becomes unity for all $\lambda_{0}$ cases. This result indicates that the bound state becomes completely composite dominant in the $B\to 0$ limit even with the four-point interaction with any strength. It is consistent with the consequence of the low-energy universality. At the same time, the decrease rate of $P_{\rm comp}$ depends on the strength of the four-point interaction $\lambda_{0}$. In other words, $\lambda_{0}$ dependence in Fig.~\ref{fig:lambda-Pcomp} expresses the model dependence of $P_{\rm comp}$ away from the $B\to 0$ limit.

\subsection{Effect of decay}
\label{subsec:decay}
Because the exotic hadrons generally have a decay width, we consider the decay effect on the compositeness in this section. To concentrate on the decay effect, here we do not include the four-point interaction and set $\lambda_{0}=0$. We effectively introduce the decay effect by letting the coupling constant $g_{0}$ be a complex number in the Hamiltonian in Eq.~\eqref{eq:H-wbs}. Because the Hamiltonian is non-Hermitian, the the eigenenergy becomes complex as
\begin{align}
E=-B-i\frac{\Gamma}{2},
\end{align}
with the decay width $\Gamma$. In the presence of the decay width $\Gamma$, the square of the coupling constant $g_{0}^{2}$ in Eq.~\eqref{eq:g02} is 
 \begin{align}
 g_{0}^{2}&=\frac{\pi^{2}}{\mu}\left(B+i\frac{\Gamma}{2}+\nu_{0}\right)\left[\Lambda-\kappa \arctan \left(\frac{\Lambda}{\kappa}\right)\right]^{-1},
 \label{eq:comp-g02} \\
 \kappa &= \sqrt{2\mu (B+i\Gamma/2)} ,
 \end{align}
which is complex for $\Gamma\neq 0$.

By definition, the compositeness $X$ and elementarity $Z$ are complex for unstable states~\cite{Kamiya:2016oao}. In fact, the compositeness $X$ in this model, obtained with Eq.~\eqref{eq:X-1ch}, is not a real number with complex $g_{0}^{2}$ and $\kappa$ for the finite $\Gamma\neq0$ case. However, we cannot interpret complex $X$ and $Z$ as probabilities as in the case of the bound state with real $X$ and $Z$. To discuss the structure of unstable states, we need to introduce other real quantities which can be interpreted as the fraction of the composite (elementary) components instead of complex $X$ ($Z$). Here we employ the quantities $\tilde{X}$ and $\tilde{Z}$ defined as
\begin{align}
\tilde{X}&=\frac{|X|}{|X|+|Z|},
\label{eq:X-tilde}\\
\tilde{Z}&=\frac{|Z|}{|X|+|Z|},
\label{eq:Z-tilde}
\end{align}
which are proposed in Ref.~\cite{Sekihara:2015gvw}. For stable states without the decay width, $\tilde{X}$ and $\tilde{Z}$ reduce to $X$ and $Z$ because $|X|=X$, $|Z|=Z$, and $X+Z=1$. It is clear that $\tilde{X}$ and $\tilde{Z}$ satisfy the sum rule:
\begin{align}
\tilde{X}+\tilde{Z}=1.
\end{align}
In addition, it follows from the definitions in Eqs.~\eqref{eq:X-tilde} and \eqref{eq:Z-tilde} that the relations $0\leq \tilde{X}\leq 1$ and $0\leq \tilde{Z}\leq 1$ hold. Therefore, we can regard $\tilde{X}$ and $\tilde{Z}$ as the probabilities of finding the composite and elementary components in a wave function instead of complex $X$ and $Z$. Hence we call $\tilde{X}$ and $\tilde{Z}$ the compositeness and the elementarity, respectively.

\begin{figure*}
\centering
\includegraphics[width=0.45\textwidth]{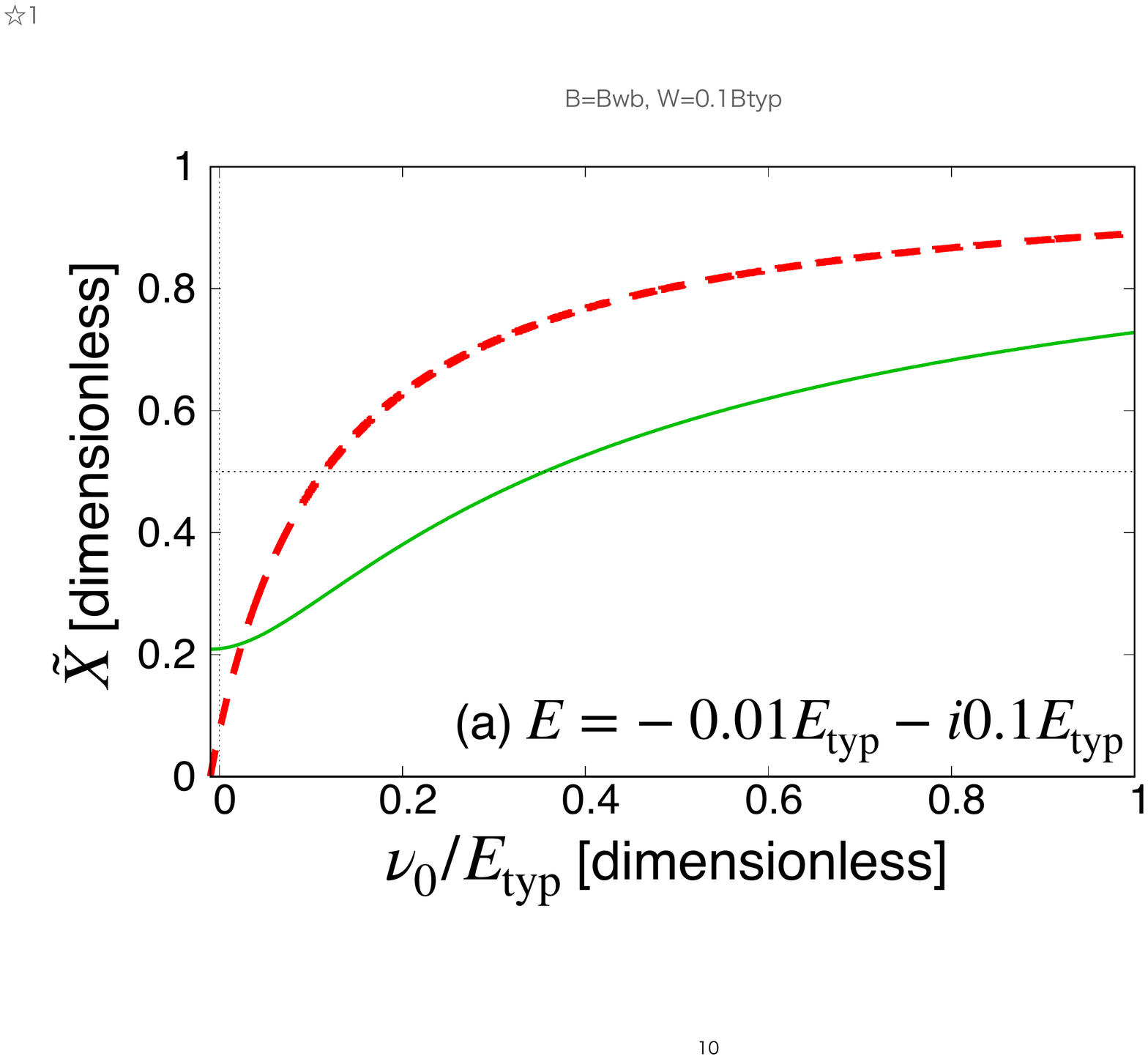} 
\includegraphics[width=0.45\textwidth]{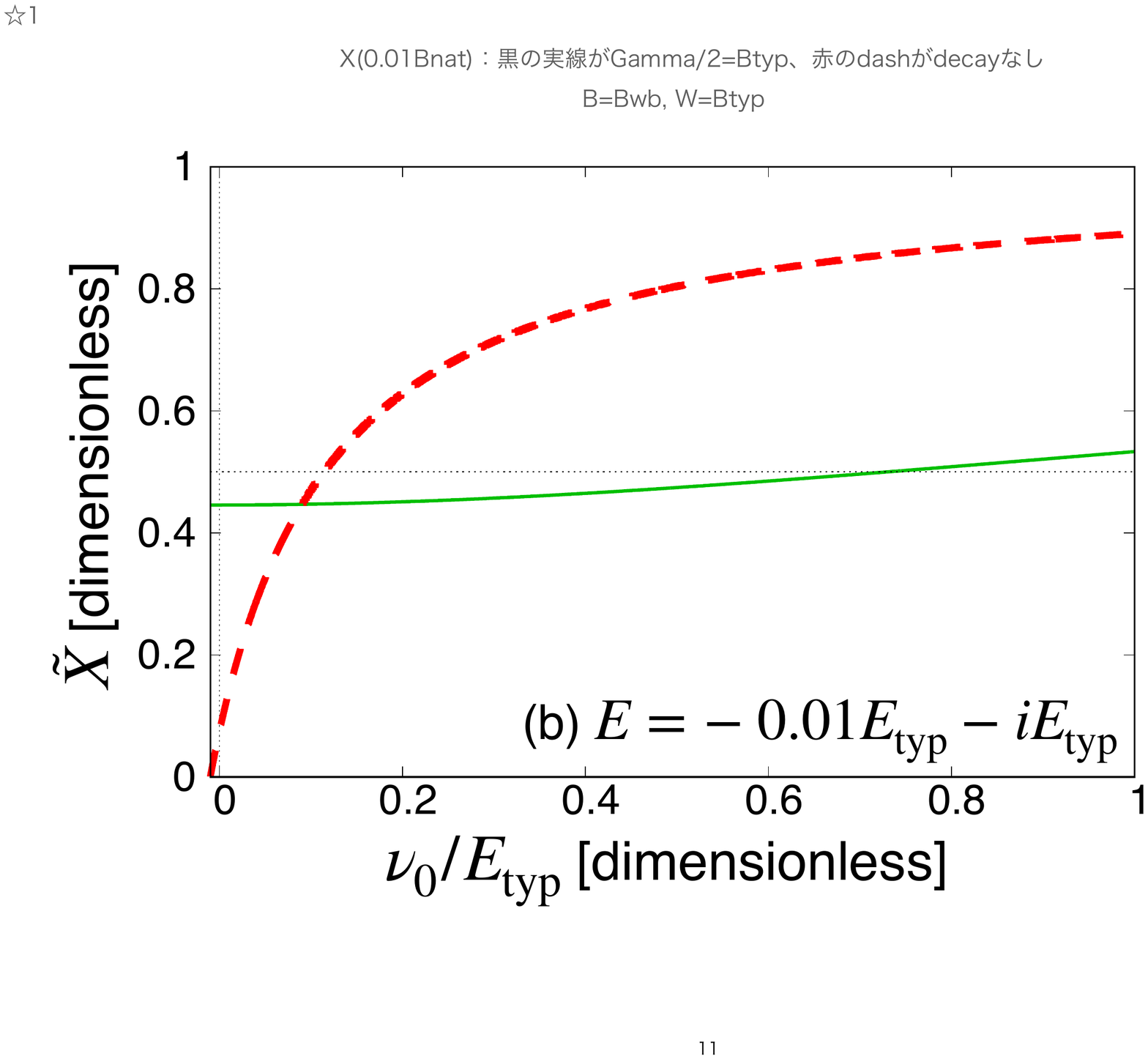} 
\includegraphics[width=0.45\textwidth]{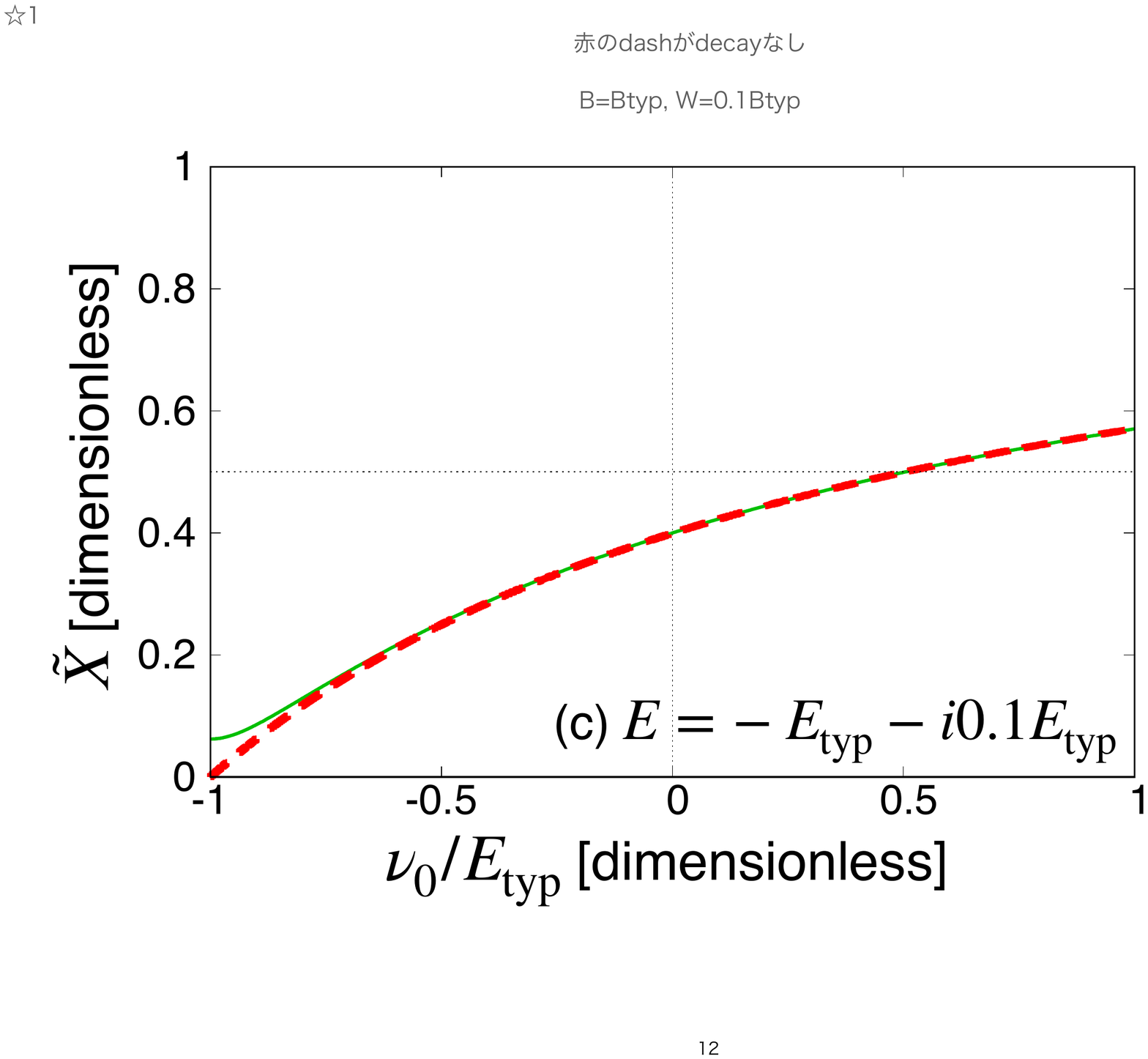} 
\includegraphics[width=0.45\textwidth]{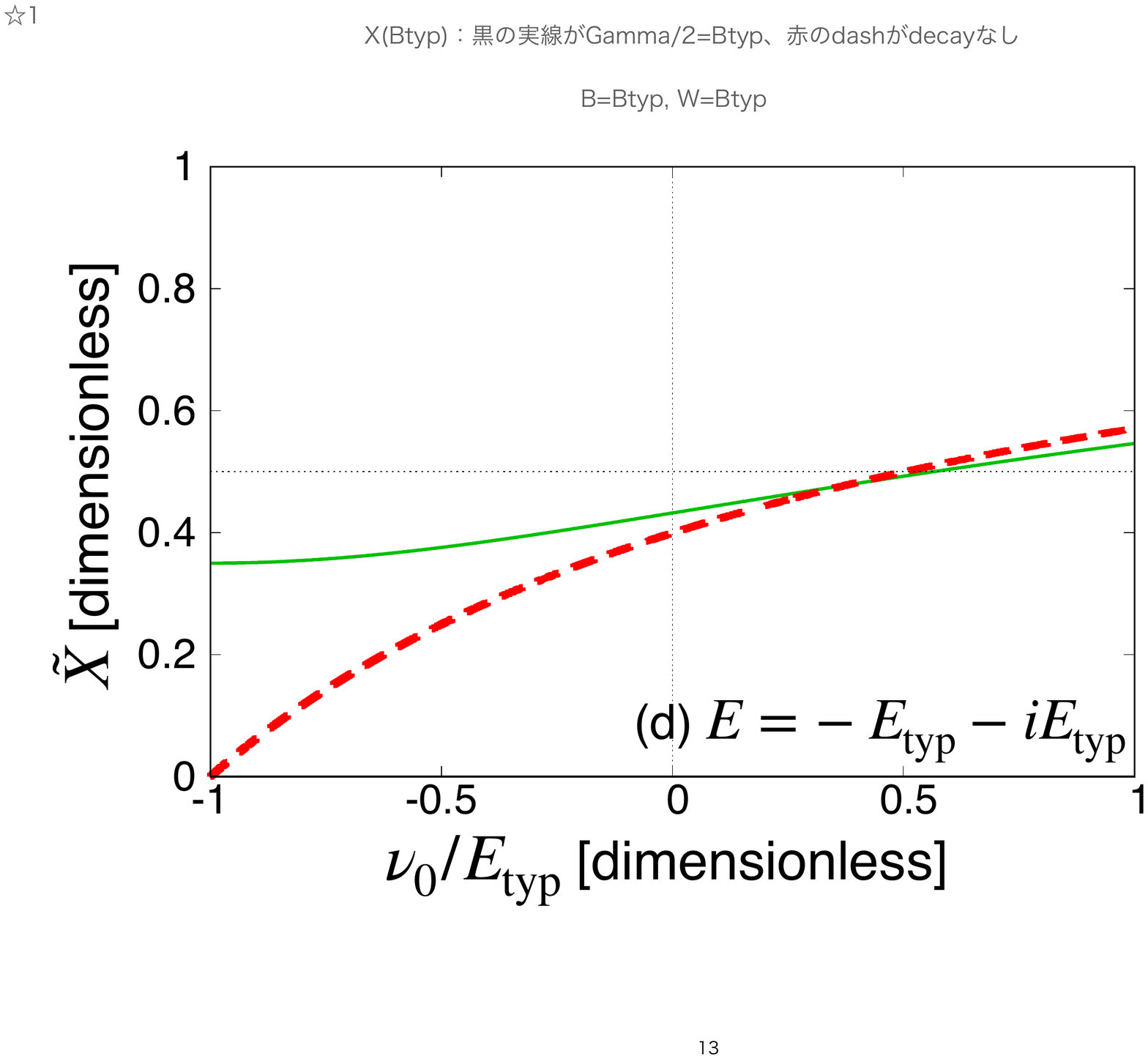} 
\caption{The compositeness $\tilde{X}$ as a function of the normalized bare state energy $\nu_{0}/E_{\rm typ}$ for $-B\leq \nu_{0}\leq E_{\rm typ}$. The solid (dashed) lines represent the results with $\Gamma\neq 0$ ($\Gamma=0$). Panel (a) corresponds to the case with $(B,\Gamma/2)=(0.01E_{\rm typ},0.1E_{\rm typ})$, (b) to $(B,\Gamma/2)=(0.01E_{\rm typ},E_{\rm typ})$, (c) to $(B,\Gamma/2)=(E_{\rm typ},0.1E_{\rm typ})$, and (d) to $(B,\Gamma/2)=(E_{\rm typ},E_{\rm typ})$.}
\label{fig:decay-X-nu}
\end{figure*}

To observe the effect of the decay, in Fig.~\ref{fig:decay-X-nu}, we plot the compositeness $\tilde{X}$ by the solid lines as a function of the normalized bare state energy $\nu_{0}/E_{\rm typ}$ for various $B$ and $\Gamma$. The panels (a) and (b) [(c) and (d)] correspond to the weak-binding (typical binding) case, and the panels (a) and (c) [(b) and (d)] represent the state with a narrow (broad) decay width. The bare state energy $\nu_{0}$ is varied in the region $-B/E_{\rm typ}\leq \nu_{0}/E_{\rm typ}\leq 1$.  For comparison, the dashed lines represent the compositeness $\tilde{X}$ for the same $B$ but with $\Gamma=0$ (same as the solid and dashed lines in Fig.~\ref{fig:X-B}).
By comparing the solid and dashed lines, we see that the effect of the decay width generally suppresses the compositeness, while $\tilde{X}$ is enhanced at small $\nu_{0}\sim -B$. Basically, the compositeness of the threshold channel decreases when the decay width is turned on because the coupling to the decay channel increases. This tendency becomes prominent especially in panel (b). The behavior of the compositeness with small $\nu_{0}$ is however governed by $\tilde{X}$ at $\nu_{0}=-B$. From Eqs.~\eqref{eq:X-1ch} and \eqref{eq:g02}, without the decay effects, the compositeness becomes zero in the $\nu_{0}\to -B$ limit because $g_{0}^{2}\to 0$. On the other hand, with a finite width $\Gamma\neq 0$, $g_{0}^{2}$ does not vanish at $\nu_{0}=-B$:
\begin{align}
g_{0}^{2}\left(-\nu_{0}+i\frac{\Gamma}{2};\nu_{0},\Lambda\right)&=\frac{\pi^{2}}{\mu}\left(-i\frac{\Gamma}{2}\right)\left[\Lambda-\kappa\arctan\left(\frac{\Lambda}{\kappa}\right)\right]^{-1}\nonumber \\ 
&\neq 0.
\label{eq:g02-Gamma}
\end{align}
From Eq.~\eqref{eq:X-1ch}, the complex compositeness $X$ becomes nonzero, and $\tilde{X}$ in Eq.~\eqref{eq:X-tilde} becomes larger than zero. This explains the enhancement of $\tilde{X}$ at $\nu_{0}\sim -B$. 

Furthermore, by comparing panels (a) and (c) with (b) and (d), we see that the $\nu_{0}$ dependence of $\tilde{X}$ becomes smaller for larger decay width. It follows from Eq.~\eqref{eq:comp-g02} that the $\nu_{0}$ dependence of $g_{0}^{2}$ is negligible for $|B+i\Gamma/2|\gg \nu_{0}$. Therefore, $\tilde{X}$ is less dependent on $\nu_{0}$, and the plot of $\tilde{X}$ becomes flat for larger $\Gamma$. For more quantitative discussion, let us analytically evaluate $\tilde{X}$ in the large decay width limit, $\Gamma\gg E_{\rm typ}$. Because $\nu_{0}$ is varied in the $-B\leq \nu_{0}\leq E_{\rm typ}$ region and the binding energy is restricted within $B\lesssim E_{\rm typ}$, the relations $\nu_{0}\ll \Gamma$ and $B\ll \Gamma$ hold under the large width limit. Furthermore, because $\kappa=\sqrt{2\mu (B+i\Gamma/2)}\sim \sqrt{i\mu \Gamma}$ and $\Gamma \gg E_{\rm typ}=\Lambda^{2}/(2\mu)$, we find $|\kappa|\gg \Lambda$ in this limit. In this case, the coupling constant $g_{0}^{2}$ in Eq.~\eqref{eq:comp-g02} behaves as 
\begin{align}
g_{0}^{2}&=\frac{3\pi^{2}\kappa^{4}}{2\mu^{2}\Lambda^{3}}+\dotsb,
\label{eq:decay-g02}
\end{align}
from the expansion of $\arctan(\Lambda/\kappa)$ for $|\kappa|\gg \Lambda$:
\begin{align}
\arctan(z)&=z-\frac{z^{3}}{3}+\mathcal{O}(z^{5})\quad (|z|\ll 1).
\end{align}
By substituting Eq.~\eqref{eq:decay-g02} into the compositeness in Eq.~\eqref{eq:X-1ch} and expanding the terms in the parenthesis by $\Lambda/\kappa$, we obtain $X$ for the large decay width limit as
\begin{align}
X&=\frac{1}{2}+\dotsb.
\end{align}
Because $Z=1/2+\dotsb$, $\tilde{X}$ is calculated as 
\begin{align}
\tilde{X}&=\frac{1}{2}+\dotsb.
\end{align}
Therefore, in the large width limit, $\tilde{X}$ approaches $1/2$ for any $\nu_0$ as expected from panels (b) and (d) in Fig.~\ref{fig:decay-X-nu}. It should, however, be noted that, in the large width limit $\Gamma\gg E_{\rm typ}$, the magnitude of the eigenenergy exceeds the applicable region of the model. Therefore, this formal limit should only be used to understand the behavior of $\tilde{X}$ with increasing $\Gamma$. 

To study the decay effect with respect to the binding energy, we compare panel (a) with (c) where the eigenstates have the same decay width. The decay effect (deviation of $\tilde{X}$ with $\Gamma\neq 0$ from that with $\Gamma=0$) in panel (a) is sizable, whereas the effect in panel (c) is almost negligible. While the half width $\Gamma/2=0.1E_{\rm typ}$ is larger than the binding energy $B=0.01E_{\rm typ}$ in panel (a), the same decay width $\Gamma/2=0.1E_{\rm typ}$ is smaller than $B=E_{\rm typ}$ in panel (c). Therefore, we conclude that the deviation of $\tilde{X}$ by the decay effect is determined by the ratio of the binding energy to the decay width.

\begin{figure}
\centering
\includegraphics[width=0.45\textwidth]{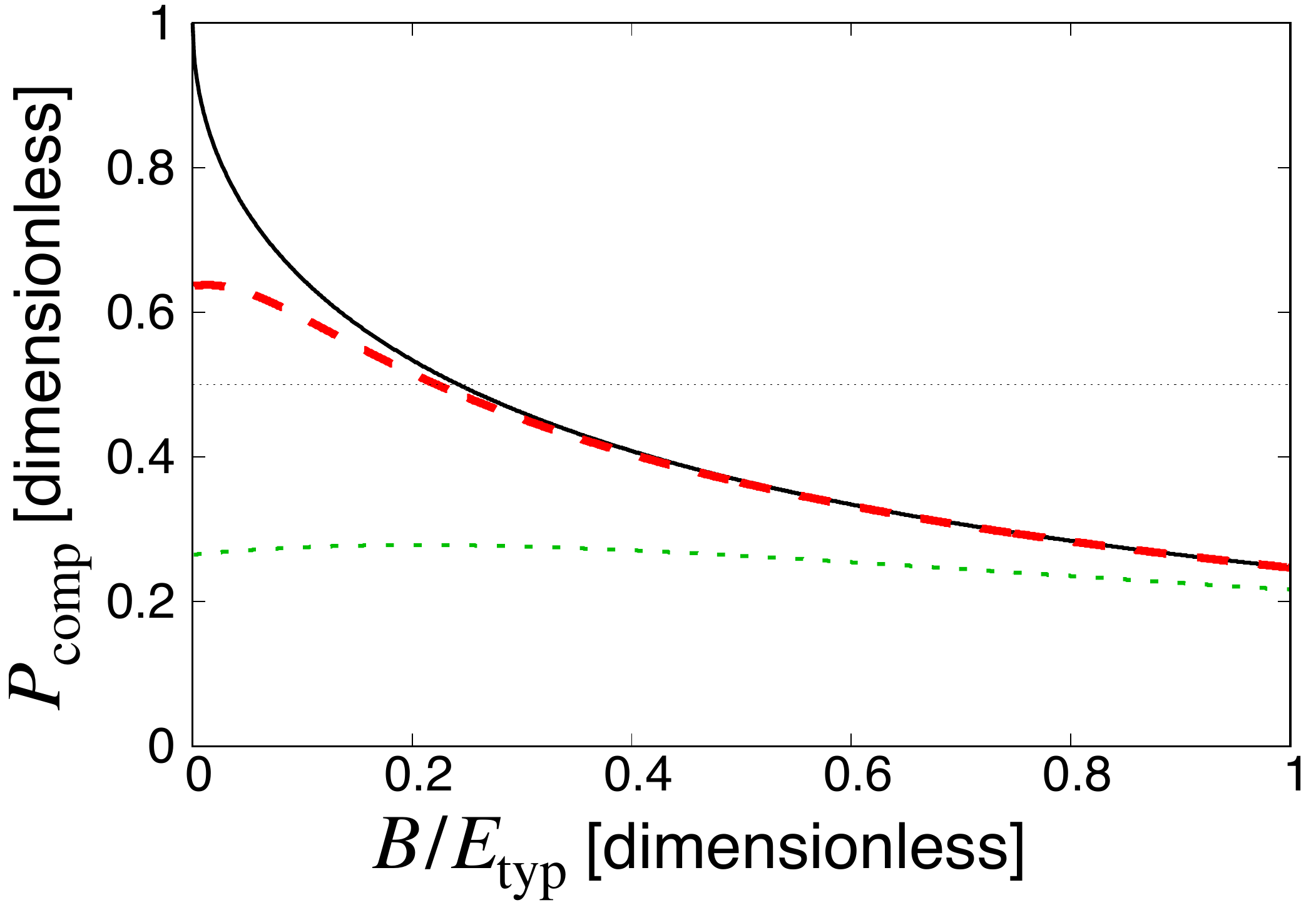}
\caption{The fraction of the composite dominant region $P_{\rm comp}$ as a function of the normalized binding energy $B/E_{\rm typ}$. The solid, dashed, and dotted lines stand for $\Gamma=0$, $\Gamma/2=0.1E_{\rm typ}$, and $\Gamma/2=E_{\rm typ}$, respectively.}
\label{fig:Pcomp-decay}
\end{figure}

To discuss the low-energy universality, we define $P_{\rm comp}$ as in Eq.~\eqref{eq:Pcomp} but with $\tilde{X}=0.5$ as the determination of $\nu_{c}$. In Fig.~\ref{fig:Pcomp-decay}, we plot $P_{\rm comp}$ as a function of the normalized binding energy $B/E_{\rm typ}$ in the presence of the decay width. The solid line stands for $P_{\rm comp}$ with $\Gamma=0$, the dashed line with $\Gamma/2=0.1E_{\rm typ}$, and the dotted line with $\Gamma/2=E_{\rm typ}$. By comparing the solid line with the dashed and dotted lines, we see that $P_{\rm comp}$ decreases when the decay width increases. This reason is understood from the $\tilde{X}$ behavior in Fig.~\ref{fig:decay-X-nu}, where $\tilde{X}$ is suppressed by introducing the decay width and $\nu_{c}$ becomes large accordingly. Therefore, the decay effect makes $P_{\rm comp}$ smaller than that for the stable states. From the $B$ dependence in Fig.~\ref{fig:Pcomp-decay}, the deviation of the dashed and dotted lines from the solid line becomes larger in the small $B$ region than that in the large $B$ region. For a fixed $\Gamma$, the ratio $\Gamma/B$ increases when the binding energy $B$ decreases. As discussed above, in this case, the compositeness is more affected by the decay. As a consequence, the change of $\nu_{c}$ is enhanced (see Fig.~\ref{fig:decay-X-nu}), and therefore the deviation of $P_{\rm comp}$ becomes large. At $B=0$, $P_{\rm comp}\neq 1$ with the finite decay width in contrast to the effect of the direct interaction in Fig.~\ref{fig:lambda-Pcomp}. With $\Gamma\neq 0$, $\kappa$ and $g_{0}^{2}$ in Eq.~\eqref{eq:comp-g02} are finite at $B=0$. Therefore, the relation $\tilde{X}=1$ does not hold with finite $\Gamma$ even in the weak-binding limit $B\to 0$, and hence $P_{\rm comp}<1$. Form the viewpoint of the universality, it is understood from the finite scattering length because the eigenenergy $E=-B-i\Gamma/2$ is nonzero for the finite $\Gamma$ even in the $B\to 0$ limit.

We note that the decay effect can be formally described in the effective field theory by introducing the decay channel in the energy region lower than the binding energy in addition to the threshold channel~\cite{Kamiya:2016oao}. In this case, we can explicitly calculate the compositeness of the decay channel. However, in this paper, we have employed the effective single-channel model with complex coupling constant $g_{0}$ in Eq.~\eqref{eq:comp-g02}, because we would like to discuss the model dependence of the compositeness by varying the bare energy $\nu_{0}$ as in Sec.~\ref{sec:wbs}. In the coupled-channels model with the explicit decay channel, a new parameter $\Delta\omega$ is introduced to express the energy difference between the threshold and decay channels as in Sec.~\ref{subsec:2ch}. To introduce the decay effect, $\Delta\omega$ is not a bare parameter but is fixed by the system. In other words, only $g_{0}$ and $\nu_{0}$ are the model parameters, which are constrained by the pole condition with a fixed eigenenergy. The complex eigenenergy of an unstable state gives two conditions from the real and imaginary parts of the pole condition. Therefore, $g_{0}$ and $\nu_{0}$ are uniquely determined in the model where the decay channel is explicitly introduced. In this case, the compositeness $X$ is not written as a function of a model parameter, and we cannot discuss the model dependence of $X$. In contrast, the number of parameters in the effective single-channel model with a complex $g_{0}$ is 3 ($\nu_{0},{\rm Re\ }g_{0},{\rm Im\ }g_{0}$). Therefore, there remains one degree of freedom even when the parameters are constrained by the pole condition. We have utilized this degree of freedom to study the $\nu_{0}$ dependence of $X$.

\subsection{Effect of channel coupling}
\label{subsec:2ch}

In the previous section, we studied the decay effect which arises from the couplings to the lower energy channel. In this section, we consider the effect of the coupling to the higher energy channel. For this purpose, we introduce the scattering of $\Psi_{1}$ and $\Psi_{2}$ (channel 2) in addition to the $\psi_{1}\psi_{2}$ scattering (channel 1) in the free Hamiltonian in Eq.~\eqref{eq:H-wbs}:
\begin{align}
\mathcal{H}_{\rm free}&=\frac{1}{2m_{1}}\nabla{\psi_{1}}^{\dag}\cdot\nabla{\psi_{1}}+\frac{1}{2m_{2}}\nabla\psi_{2}^{\dag}\cdot\nabla{\psi_{2}}\nonumber \\
&\quad+\frac{1}{2M_{1}}\nabla{\Psi_{1}}^{\dag}\cdot\nabla{\Psi_{1}}+\frac{1}{2M_{2}}\nabla\Psi_{2}^{\dag}\cdot\nabla{\Psi_{2}}\nonumber \\
&\quad +\frac{1}{2M}\nabla\phi^{\dag}\cdot\nabla\phi+\omega_{1}\Psi_{1}^{\dag}\Psi_{1}+\omega_{2}\Psi_{2}^{\dag}\Psi_{2}+\nu_{0}\phi^{\dag}\phi.
\label{eq:H-free-3}
\end{align}
where $M_{1}$ and $M_{2}$ are the masses of $\Psi_{1}$ and $\Psi_{2}$, and $\omega_{1}$ and $\omega_{2}$ are the energies of $\Psi_{1}$ and $\Psi_{2}$ measured from the $\psi_{1}\psi_{2}$ threshold. $\Delta \omega=\omega_{1}+\omega_{2}>0$ is the threshold energy difference between channels 1 and 2. For the transition from channel 1 to channel 2, here we introduce the coupling of channel 2 and the bare state $\phi$. We employ the same coupling constant $g_{0}$ as that for channel 1 and $\phi$. The interaction Hamiltonian leads to
\begin{align}
\mathcal{H}_{\rm int}&=g_{0}(\phi^{\dagger}\psi_{1}\psi_{2}+\psi^{\dagger}_{1}\psi^{\dagger}_{2}\phi+\phi^{\dagger}\Psi_{1}\Psi_{2}+\Psi^{\dagger}_{1}\Psi^{\dagger}_{2}\phi).
\label{eq:H-int-3}
\end{align}
 
We now consider the on-shell T-matrix $T_{\rm on}(k)$ of the coupled-channels scatterings. As in the single-channel case in Sec.~\ref{subsec:eft-2}, the scatterings occur through the effective interaction with the bare state $\phi$ exchange. In the coupled-channel scattering, $T_{\rm on}(k)$, the effective interaction $V(k)$ and the loop function $G(k)$ are expressed by the matrices in the channel space. In this model, the on-shell T-matrix is given by
\begin{align}
T_{\rm on}(k_{1})&=V(k_{1})+V(k_{1})G(k_{1})T_{\rm on}(k_{1}),\\
V(k_{1})&=\begin{pmatrix}
v(k_{1}) & v(k_{1})\\
v(k_{1}) & v(k_{1})
\end{pmatrix}, 
\label{eq:V-3}\\
G(k_{1})&=\begin{pmatrix}
G_{1}(k_{1}) & 0\\
0 & G_{2}(k_{2}(k_{1}))
\end{pmatrix}.
\end{align}
Here each component of $V(k)$ and $G(k)$ is 
\begin{align}
v(k_{1})&=\frac{g_{0}^{2}}{\frac{k_{1}^{2}}{2\mu_{1}}-\nu_{0}},
\label{eq:v-3}\\
G_{i}(k_{i})
&=-\frac{\mu_{i}}{\pi^{2}}\left[\Lambda+ik_{i}\arctan\left(-\frac{\Lambda}{ik_{i}}\right)\right],
\label{eq:G1}
\end{align}
where the momentum of each channel at the energy $E$ is
\begin{align}
k_{1}&=\sqrt{2\mu_{1}E},\\
k_{2}(k_{1})&=\sqrt{2\mu_{2}(E-\Delta\omega)}=\sqrt{\frac{\mu_{2}}{\mu_{1}}k_{1}^{2}-2\mu_{2}\Delta \omega}, 
\end{align}
with $\mu_{1}=(1/m_{1}+1/m_{2})^{-1}$ and $\mu_{2}=(1/M_{1}+1/M_{2})^{-1}$. 

As before, we assume that there is a bound state. The bound state condition for the coupled-channels scattering is given by $\det(1-GV)=0$. This leads to
\begin{align}
E-\nu_{0}-g_{0}^{2}[G_{1}(k_{1})+G_{2}(k_{2})]=0,
\end{align}
with $E=-B$.
By solving this condition for $g_{0}^{2}$, we obtain the expression of $g_{0}^{2}$ as
\begin{align}
g_{0}^{2}(B;\nu_{0},\Lambda)=-\frac{B+\nu_{0}}{G_{1}(i\kappa_{1})+G_{2}(i\kappa_{2})},
\end{align}
with $\kappa_{1}=\sqrt{2\mu_{1}B}$ and $\kappa_{2}=\sqrt{2\mu_{2}(B+\Delta\omega)}$

In the coupled-channel scattering, the compositeness is defined for each channel as $X_{1}$ and $X_{2}$. $X_{i}$ is interpreted as the probability of finding channel $i$ composite state in the bound state. As in the single-channel case, the compositenesses $X_{1}$, $X_{2}$ and the elementarity $Z=1-X_{1}-X_{2}$ are calculated from the effective interaction in Eq.~\eqref{eq:v-3} and the loop functions in Eq.~\eqref{eq:G1}. As discussed in Ref.~\cite{Kamiya:2016oao}, the expression of $X_{1}$ is obtained by replacing $G\to G_{1}$ and $V^{-1}\to [v_{\rm eff}]^{-1}$ in Eq.~\eqref{eq:X-1ch}, where $v_{\rm eff}$ is the effective interaction in channel~1 obtained by eliminating the bare state and channel~2. In the present model, the effective interaction is~\cite{Kamiya:2016oao}
\begin{align}
[v_{\rm eff}]^{-1}(k_{1})&=\frac{1-G_{2}(k_{2})v(k_{1})}{[1-G_{2}(k_{2})v(k_{1})]v(k_{1})+G_{2}(k_{2})v^{2}(k_{1})}\nonumber \\
&=v^{-1}(k_{1})-G_{2}(k_{2}).
\end{align}
Then the compositenesses $X_{1}$ and $X_{2}$ are 
\begin{align}
X_{1}&=\frac{G'_{1}(i\kappa_{1})}{G'_{1}(i\kappa_{1})+G'_{2}(i\kappa_{2})-[v^{-1}]'},
\label{eq:X1}\\
X_{2}&=\frac{G'_{2}(i\kappa_{2})}{G'_{1}(i\kappa_{1})+G'_{2}(i\kappa_{2})-[v^{-1}]'},
\label{eq:X2}
\end{align}
where $\kappa_{1}=\sqrt{2\mu_{1}B}$, $\kappa_{2}=\sqrt{2\mu_{2}(B+\Delta\omega)}$, and the derivatives of $v^{-1}$ and the loop functions $G_{1},G_{2}$ are given by
\begin{align}
[v^{-1}]'&=\frac{1}{g_{0}^{2}},
\label{eq:v-inv-prime}\\
G'_{i}(i\kappa_{i})&=-\frac{\mu_{i}^{2}}{\pi^{2}\kappa_{i}}\left[\arctan\left(\frac{\Lambda}{\kappa_{i}}\right)-\frac{\frac{\Lambda}{\kappa_{i}}}{1+\left(\frac{\Lambda}{\kappa_{i}}\right)^{2}}\right],
\label{eq:G-prime}
\end{align}
with $i=1$ and $2$.

For the numerical calculation, we can choose arbitrarily $\mu_{1,2}$ and $\Delta \omega$ by adjusting $m_{1,2}$, $M_{1,2}$, and $\omega_{1,2}$. With the dimensionless parameters, the result only depends on the ratio of $\mu_{1}$ and $\mu_{2}$. In this section, to focus on the $\Delta \omega$ dependence, we assume $\mu_{1}=\mu_{2}$. 

\begin{figure*}
\centering
\includegraphics[width=0.45\textwidth]{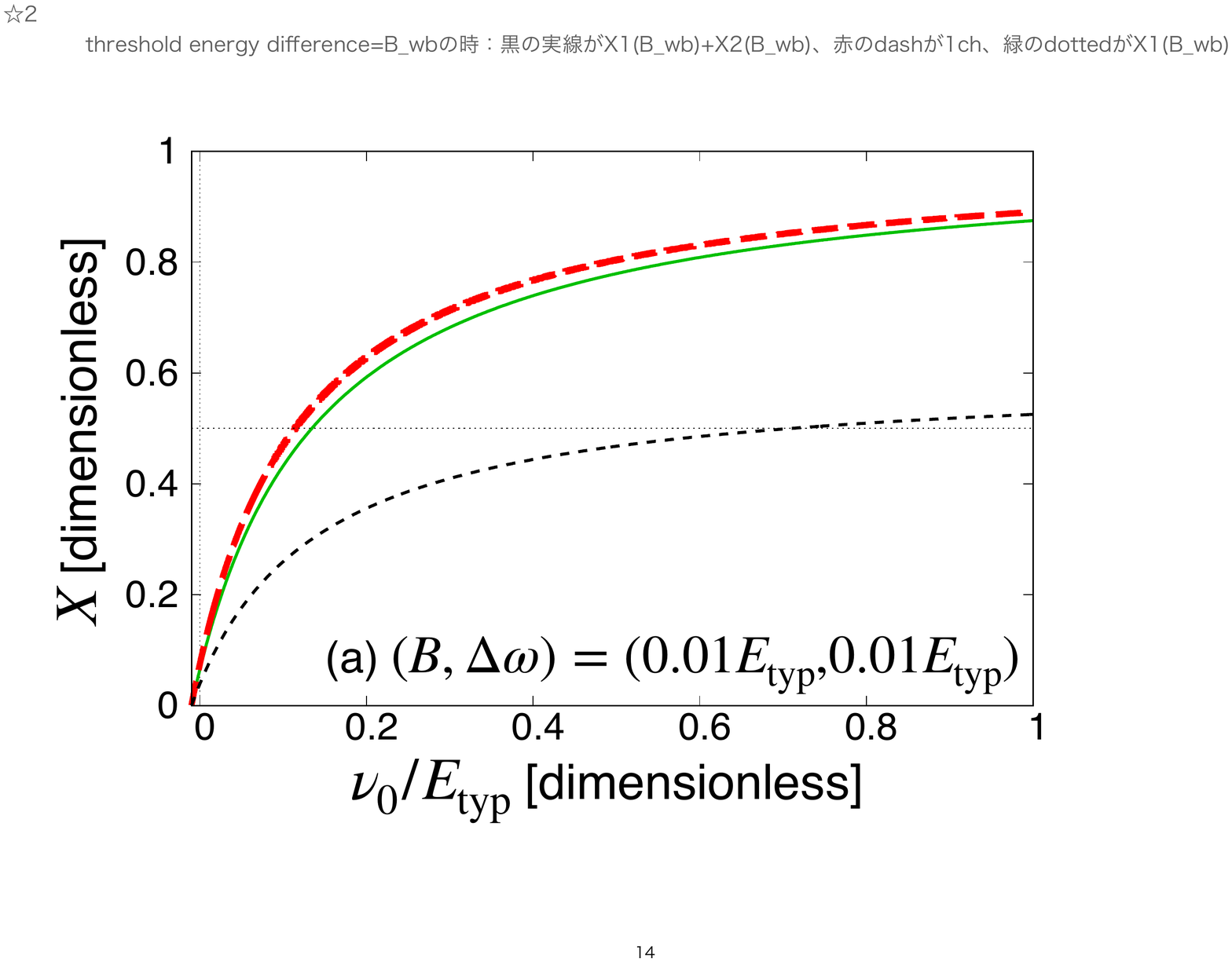} 
\includegraphics[width=0.45\textwidth]{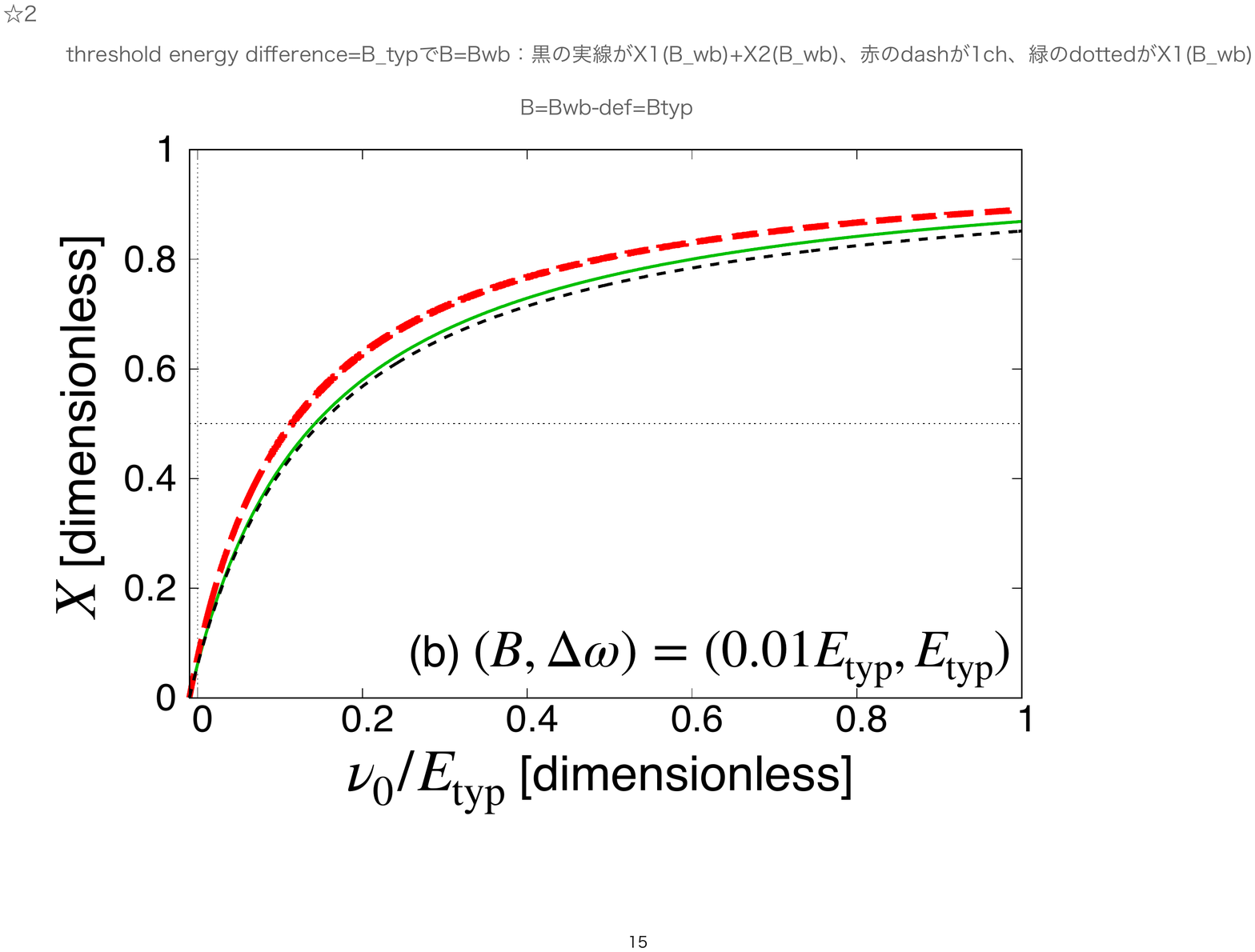} 
\includegraphics[width=0.45\textwidth]{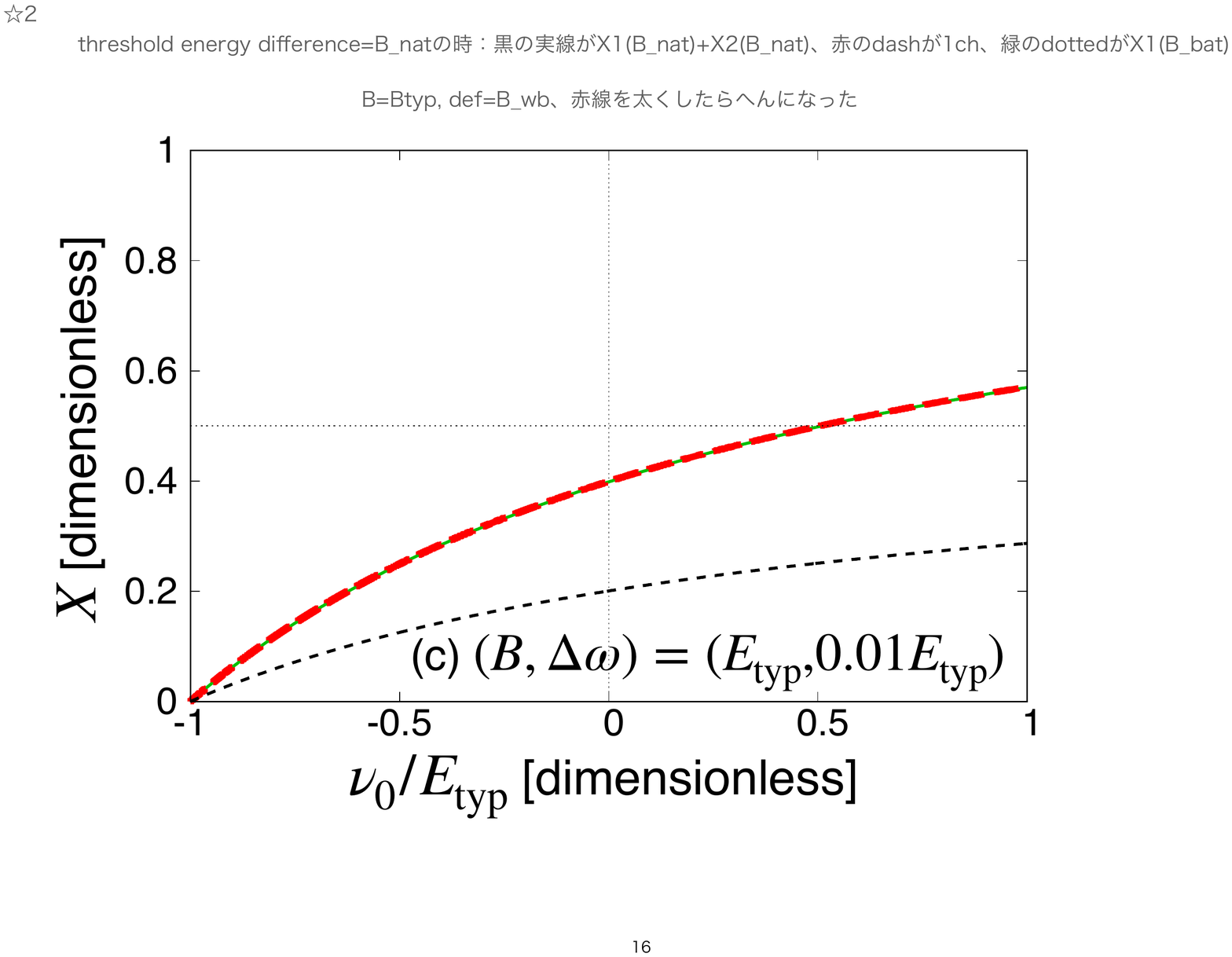} 
\includegraphics[width=0.45\textwidth]{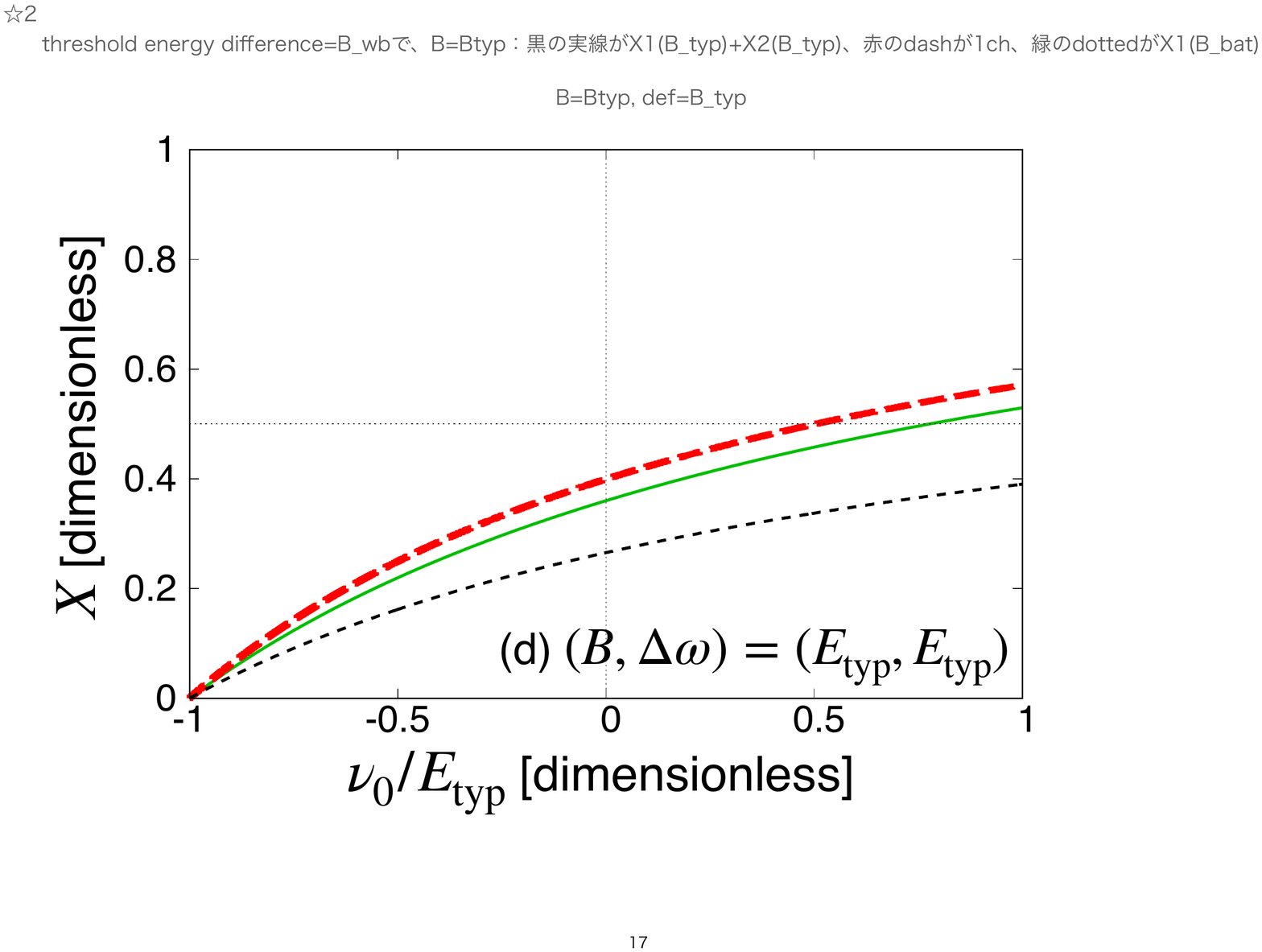} 
\caption{The compositeness as a function of the normalized bare state energy $\nu_{0}/E_{\rm typ}$ for $-B\leq \nu_{0}\leq E_{\rm typ}$ at fixed binding energy and the threshold energy difference $(B,\Delta \omega)=(0.01E_{\rm typ},0.01E_{\rm typ})$ [panel (a)], $(B,\Delta \omega)=(0.01E_{\rm typ},E_{\rm typ})$ [panel (b)], $(B,\Delta \omega)=(E_{\rm typ},0.01E_{\rm typ})$ [panel (c)], and $(B,\Delta \omega)=(E_{\rm typ},E_{\rm typ})$ [panel (d)]. The solid lines represent for $X_{1}+X_{2}$, the dotted lines $X_{1}$, and the
dashed lines the compositeness in the the single-channel case.
}
\label{fig:2chX-nu}
\end{figure*}

To quantitatively study the contribution of the coupled channel, we plot the compositeness as a function of the normalized bare state energy $\nu_{0}/E_{\rm typ}$ in Fig.~\ref{fig:2chX-nu}. The solid lines stand for $X_{1}+X_{2}$ and the dotted lines for $X_{1}$. Therefore, the difference between the solid and dotted lines corresponds to $X_{2}$. To see the coupled-channels effect to the compositeness, we plot $X$ in Eq.~\eqref{eq:X-1ch} for single-channel case with same $B$ and $\nu_{0}$ by the dashed lines. Panels (a) and (b) [(c) and (d)] correspond to the weak-binding (typical binding) case, and panels (a) and (c) [(b) and (d)] show the results with small (large) threshold energy difference $\Delta \omega$. By comparing panels (a) with (b) and (c) with (d), we see that $X_{2}$ becomes smaller for larger threshold energy difference $\Delta\omega$. This is analytically explained by the behavior of $X_{2}$ in Eq.~\eqref{eq:X2} in the large $\Delta\omega$ limit. When $\Delta\omega\to \infty$, $\kappa_{2}$ also goes to infinity. This induces that $G'_{2}\to 0$ in Eq.~\eqref{eq:G-prime} and $X_{2}$ becomes zero in Eq.~\eqref{eq:X2}. Intuitively, this is because the channel 2 contribution vanishes when the threshold is infinitely far away. The limit $\Delta\omega\to \infty$ is considered only as the formal limit to understand the behavior of $X_{2}$ with large $\Delta\omega$. We note that larger $\Delta \omega$ than $E_{\rm typ}$ exceeds the applicable region of the model. 

In the opposite limit $\Delta\omega\to 0$, we can also analytically show that $X_{1}=X_{2}$, because $\kappa_{1}=\kappa_{2}$ and then $G_{1}=G_{2}$ under the assumption of this calculation with $\mu_{1}=\mu_{2}$, so Eq.~\eqref{eq:X2} becomes identical with Eq.~\eqref{eq:X1}. This is because the physical bound state couples to both the channels with an equal weight. This behavior is reflected in panel (c), where $\Delta\omega$ is negligibly smaller than $B$ and the dotted line indicates about half of the solid line.
We note that the ratio $X_{1}/X_{2}$ in the $\Delta\omega\to 0$ limit depends on the bare coupling strengths in channels 1 and 2. In this work, we obtain $X_{1}/X_{2}=1$ because the common coupling constant $g_{0}$ to both the channels is adopted in the interaction Lagrangian in Eq.~\eqref{eq:H-int-3}. With different coupling strengths for channels 1 and 2, we obtain the ratio $X_{1}/X_{2}\neq 1$ in the $\Delta\omega\to 0$ limit. 

It is also observed in all panels in Fig.~\ref{fig:2chX-nu} that the sum $X_{1}+X_{2}$ (the solid line) is close to $X$ in the single-channel model (the dashed line). In our coupled-channels model, the bound state is formed by the dressing of the bare state through the coupling to the scattering states. The dressing induces the two-body composite component to the eigenstate and increases the compositeness. From a fixed bare state energy $\nu_{0}$, we need the same amount of the dressing to obtain the bound state at $E=-B$, irrespective of the number of coupled channels. In the coupled-channels model, channels 1 and 2 work cooperatively to achieve the dressing equivalent to that in the single-channel model. In other words, the compositeness $X_{1}+X_{2}\sim X$ represents the total amount needed to dress the bare state to the bound state.

\begin{figure}
\centering
\includegraphics[width=0.5\textwidth]{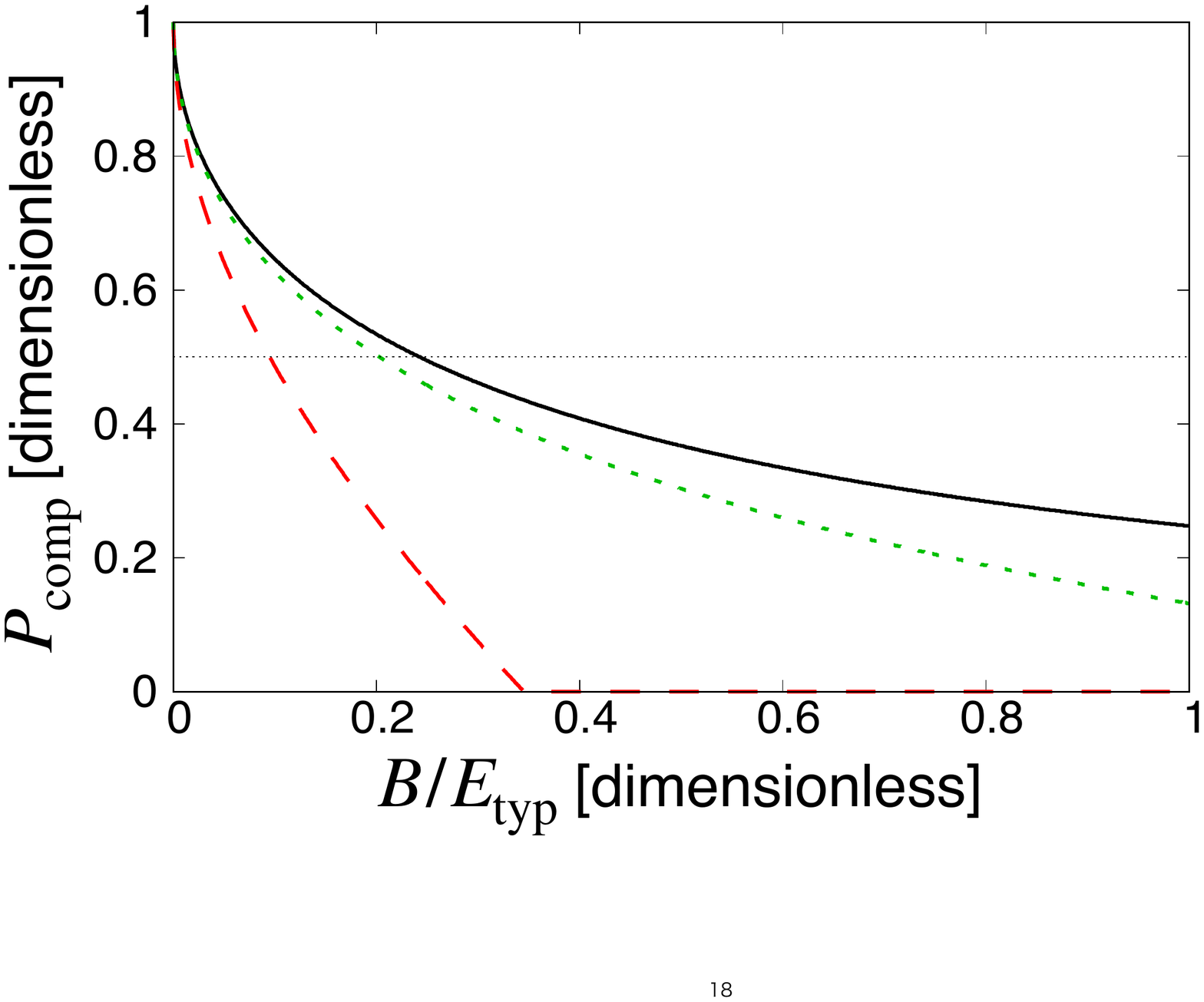}
\caption{The fraction of the threshold channel composite dominant region $P_{\rm comp}$ as a function of the normalized binding energy $B/E_{\rm typ}$ for the fixed threshold energy difference. The solid line represents to the single-channel case, the dotted line to $\Delta\omega=10E_{\rm typ}$, and the dashed line to $\Delta\omega=E_{\rm typ}$.}
\label{fig:Pcomp-omega}
\end{figure}

For the multi channel case, the low-energy universality indicates that the bound state is completely dominated by the threshold channel, namely, $X_{1}=1$, $X_{2}=0$, and $Z=0$ in the $B\to 0$ limit. To focus on the dominance of $X_{1}$, we define $P_{\rm comp}$ in Eq.~\eqref{eq:Pcomp} with $\nu_{c}$, which gives $X_{1}=0.5$ as the probability of finding a model with the $\psi_{1}\psi_{2}$ composite dominant state. In Fig.~\ref{fig:Pcomp-omega}, we plot $P_{\rm comp}$ as a function of the normalized binding energy $B/E_{\rm typ}$ for $\Delta\omega=E_{\rm typ}$ (dashed line), $\Delta\omega=10E_{\rm typ}$ (dotted line), and the single-channel case (solid line). By comparing the three lines, we find that $P_{\rm comp}$ in the coupled-channels case is suppressed compared to that in the single-channel case at the same $B$, and the suppression becomes larger for smaller $\Delta \omega$. The reason for this is seen as the change of $\nu_{c}$ in panels (a) and (b) in Fig.~\ref{fig:2chX-nu}; $\nu_{c}/E_{\rm typ}= 0.15$ for $\Delta\omega=E_{\rm typ}$ [panel (b)] changes to $\nu_{c}/E_{\rm typ}= 0.71$ for $\Delta\omega=0.01E_{\rm typ}$ [panel (a)] so that the fraction of the composite dominant region decreases. In Fig.~\ref{fig:Pcomp-omega}, the dashed line becomes zero in the region $B/E_{\rm typ}\geq 0.35$, where the channel 1 compositeness $X_{1}$ is always smaller than 0.5 and there is no $X_{1}$ dominant region [see panel (d) in Fig.~\ref{fig:2chX-nu}]. At $B=0$ in Fig.~\ref{fig:Pcomp-omega}, $P_{\rm comp}$ becomes unity even with the coupled-channels effect with finite $\Delta \omega$. For arbitrary $\Delta \omega \neq 0$, one can always consider the small binding energy $B$ such that $B\ll \Delta \omega$. In this case, the bound state decouples from channel 2, and $X_{2}$ becomes zero as discussed above in the $\Delta\omega\to\infty$ limit. At the same time, the bound state is completely dominated by the composite component of the threshold channel, $X_{1}\to 1$. This is consistent with the consequence of the low-energy universality.

\section{Application to $T_{cc}$ and $X(3872)$}
\label{sec:apply}
Based on the properties of the near-threshold states discussed so far in general cases, we now consider the application to hadron physics. As prominent examples of  weakly bound exotic hadrons, we discuss the nature of $T_{cc}$ and $X(3872)$ by calculating the compositeness with the effective field theory. As mentioned in the Introduction (see Fig.~\ref{fig:Tcc-X3872}), $T_{cc}$ is observed slightly below the $D^{0}D^{*+}$ threshold, and the coupled channel of the isospin partner $D^{*0}D^{+}$ exists above the threshold channel. Similarly, $X(3872)$ is the weakly bound state near the $D^{0}\bar{D}^{*0}$ threshold, and couples to the $D^{+}D^{*-}$ channel above the threshold. Both the states decay through the strong interaction. Therefore, to analyze the structure of $T_{cc}$ and $X(3872)$, we introduce both contributions of the decay and the channel coupling discussed in Sec.~\ref{sec:effects}. 

As mentioned in Sec.~\ref{subsec:decay}, for an unstable state, we need to introduce $\tilde{X}$ in Eq.~\eqref{eq:X-tilde} as the compositeness because the complex compositeness is not interpreted as a probability. In the coupled-channels case, we define $X_{1}$ and $X_{2}$ as the compositenesses of the threshold and coupled channels, respectively, as in Sec.~\ref{subsec:2ch}. To take into account both the decay and the coupled-channels contributions, we employ $\tilde{X}_{1}$ and $\tilde{X}_{2}$ proposed in Ref.~\cite{Sekihara:2015gvw}:
\begin{align}
\tilde{X}_{j}&=\frac{|X_{j}|}{\sum_{j}|X_{j}|+|Z|},\quad (j=1,2).
\label{eq:X-tilde-multi}
\end{align}
$\tilde{X}_{1}$ and $\tilde{X}_{2}$ can be interpreted as the probabilities of finding the threshold and the coupled channels components, respectively. 

For the numerical calculation, the masses of the $D$ mesons are taken from the Particle Data Group (PDG)~\cite{ParticleDataGroup:2020ssz}. We employ the binding energy and the decay width of $T_{cc}$ from the pole parameters in Ref.~\cite{LHCb:2021auc} and those of $X(3872)$ from PDG~\cite{ParticleDataGroup:2020ssz}:
\begin{align}
T_{cc}&: E=-0.36^{+0.044}_{-0.040}-i0.024^{+0.001}_{-0.007}\ {\rm MeV}, \label{eq:Tcc-energy} \\
X(3872)&: E=-0.04\pm 0.06-i0.595\pm 0.105\ {\rm MeV}. \label{eq:X3872-energy}
\end{align}
We use the cutoff $\Lambda = m_{\pi}=140$ MeV because $\pi$ can be exchanged between the $D$ mesons. In this case, the typical binding energy scales are obtained as $E_{\rm typ}=10.13$~MeV and $E_{\rm typ}=10.14$~MeV for the $T_{cc}$ and $X(3872)$ systems, respectively. 
The coupled-channels and decay effects are characterized by the threshold energy difference $\Delta \omega$ and the decay width $\Gamma$. In the $T_{cc}$ case, the energy difference between the threshold channel and the coupled channel is $\Delta \omega=1.41$ MeV, and the central value of the decay width is $\Gamma=0.048$ MeV. In the $X(3872)$ case, the energy difference is $\Delta \omega=8.23$ MeV, and the decay width is $\Gamma=1.19$ MeV. In this way, we have smaller threshold energy difference $\Delta\omega$ and decay width $\Gamma$ for $T_{cc}$, and larger $\Delta\omega$ and $\Gamma$ for $X(3872)$, as shown in Fig.~\ref{fig:Tcc-X3872}. 

\begin{figure*}
\centering
\includegraphics[width=0.45\textwidth]{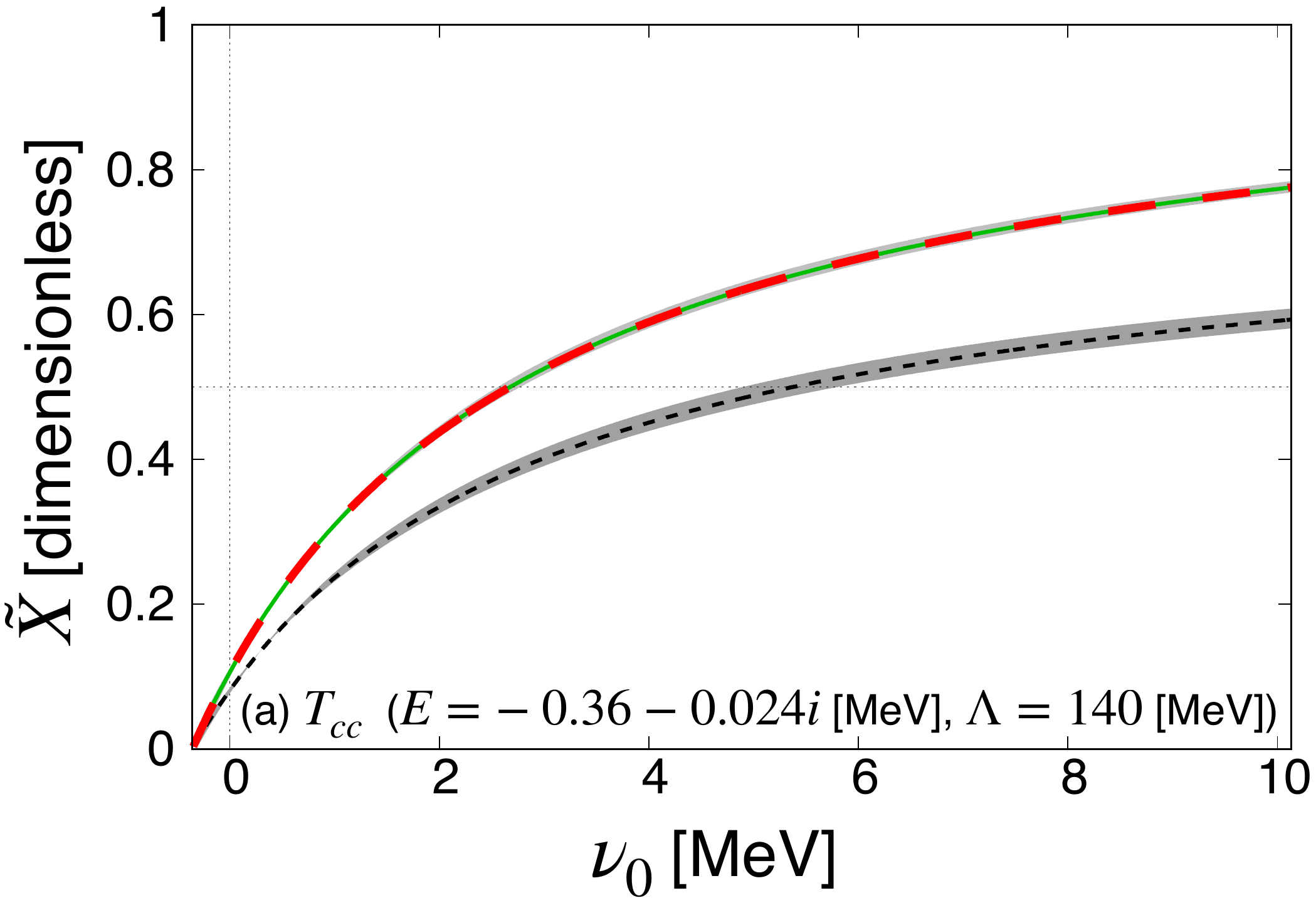}
\includegraphics[width=0.45\textwidth]{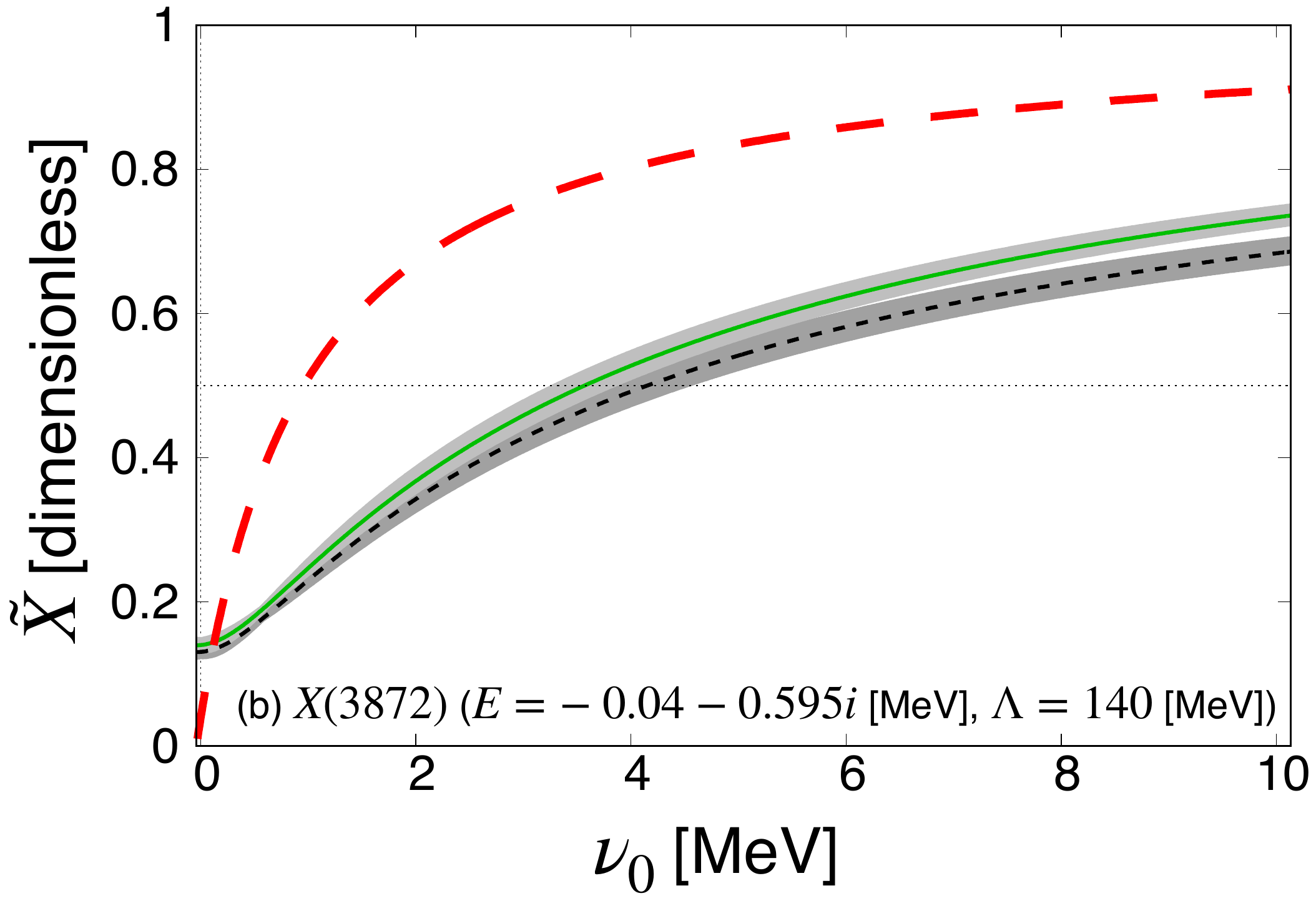}
\caption{The compositeness $\tilde{X}$ as a function of the bare state energy $\nu_{0}$. Panel (a) [(b)] shows the result of $T_{cc}$ [$X(3872)$]. The solid lines stand for the sum of the compositeness of threshold and coupled channels, $\tilde{X}_{1}+\tilde{X}_{2}$, the dotted lines show $\tilde{X}_{1}$, and the dashed lines show $\tilde{X}_{1}+\tilde{X}_{2}$ with setting $\Gamma=0$. The cutoff is fixed as $\Lambda=140$ MeV.}
\label{fig:apply}
\end{figure*}

\begin{figure*}
\centering
\includegraphics[width=0.45\textwidth]{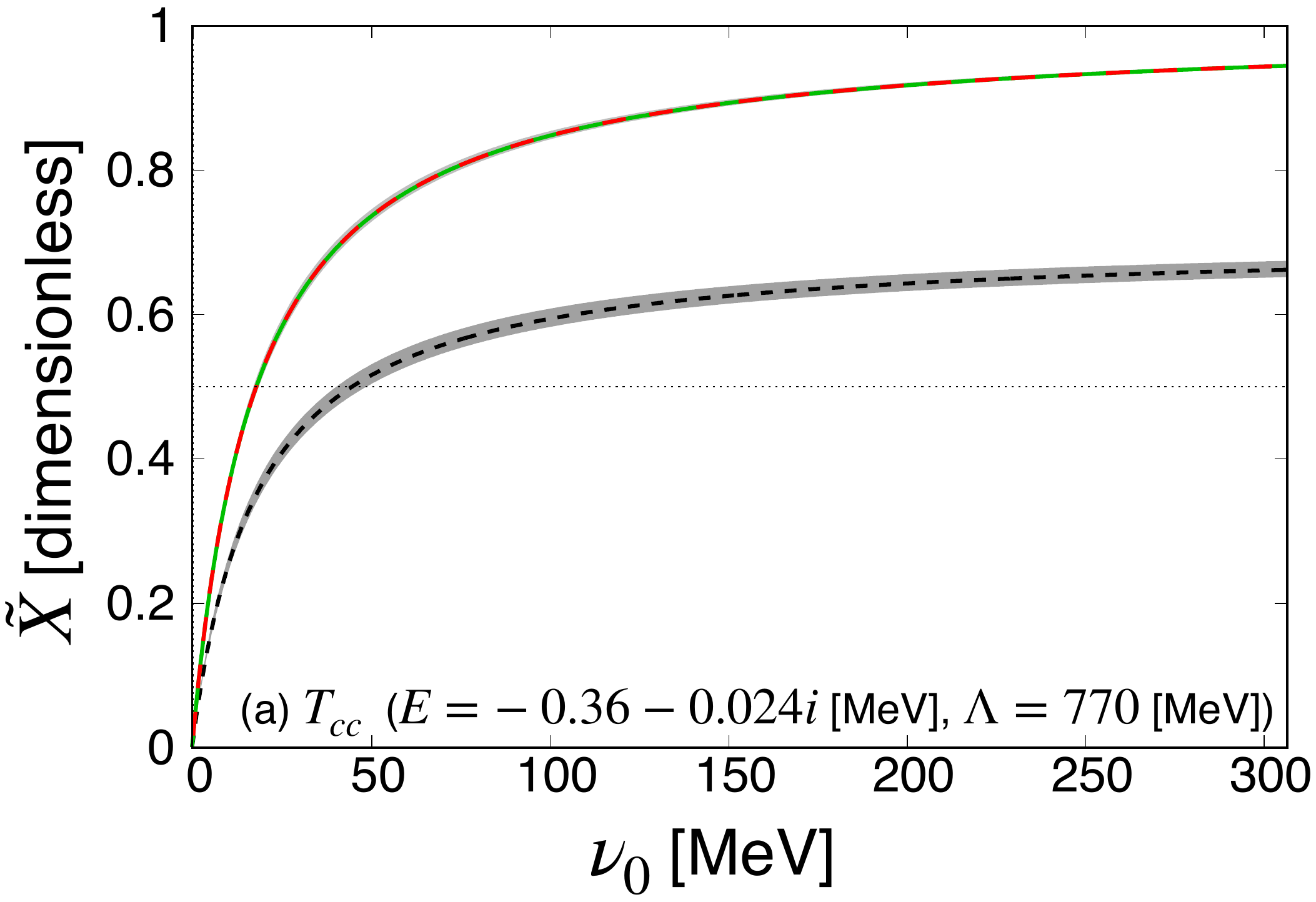}
\includegraphics[width=0.45\textwidth]{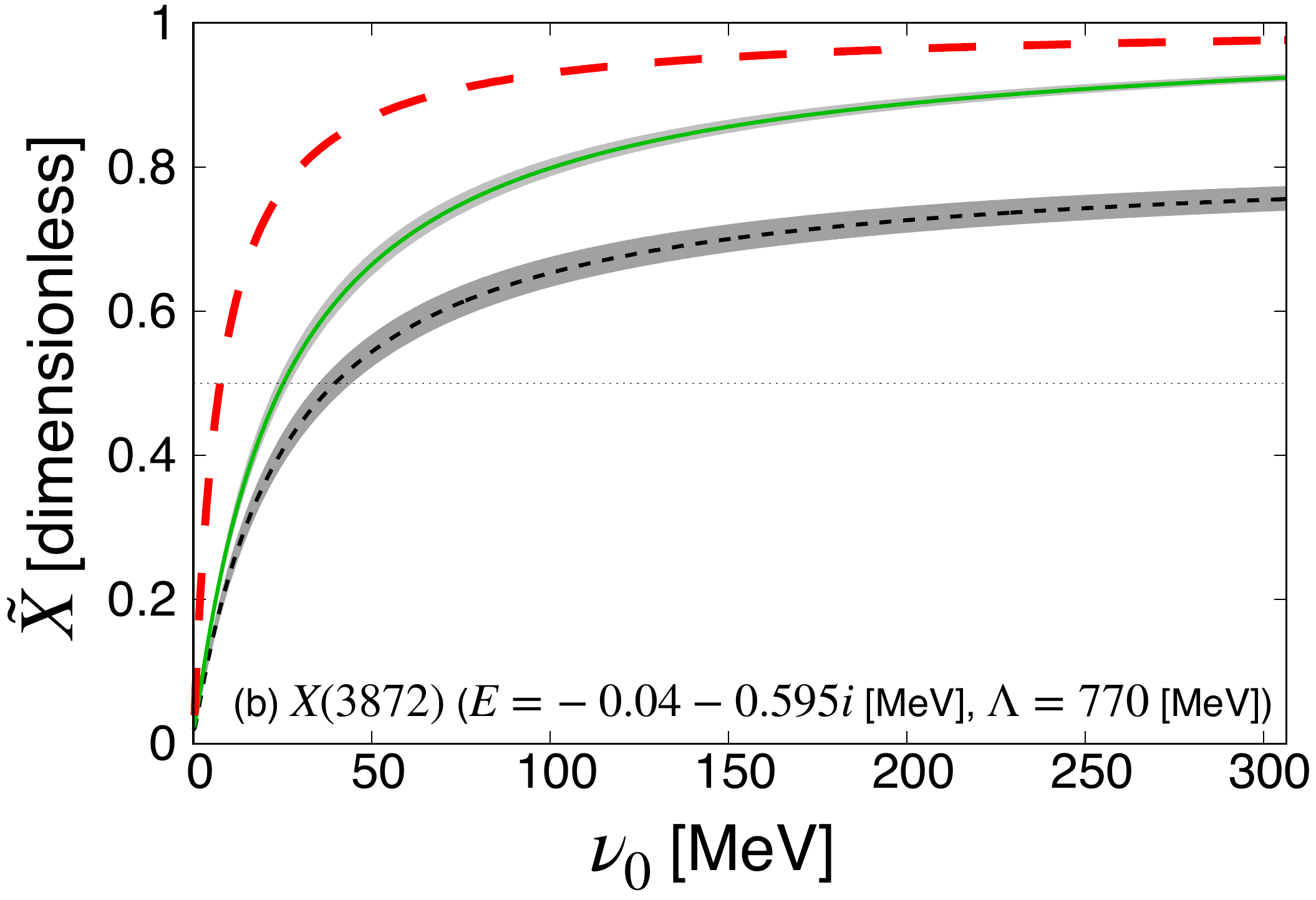}
\caption{Same as Fig.~\ref{fig:apply}, but the cutoff is fixed as $\Lambda=770$ MeV.}
\label{fig:apply-770}
\end{figure*}

In Fig.~\ref{fig:apply}, we plot the compositenesses of $T_{cc}$ [panel (a)] and $X(3872)$ [panel (b)] as a function of the bare state energy $\nu_{0}$. The solid lines represent $\tilde{X}_{1}+\tilde{X}_{2}$, and the dotted lines $\tilde{X}_{1}$. For comparison, we show by the dashed lines $\tilde{X}_{1}+\tilde{X}_{2}$ with artificially setting $\Gamma=0$. In panel (a), the solid line almost overlaps with the dashed line and the deviation of $\tilde{X}_{1}+\tilde{X}_{2}$ due to the decay width is too small to observe. This is because $\tilde{X}_{1}+\tilde{X}_{2}$ of $T_{cc}$ does not change when the narrow decay width ($0.048$ MeV) is turned on. In contrast, in panel (b), we find a sizable deviation due to the decay effect. Although the decay width of $X(3872)$ ($1.19$ MeV) is small in hadron physics, it is nevertheless larger than the binding energy ($0.04$ MeV) and the magnitude of the decay effect reflects the ratio of the binding energy to the decay width as we discussed in Sec.~\ref{subsec:decay}. Next, we consider the effect of the channel coupling, indicated by the difference between the solid and dotted lines. In Fig.~\ref{fig:apply}, we see that the difference between those lines is larger in panel (a) than in panel (b). In other words, the coupled-channels contribution $\tilde{X}_{2}$ of $X(3872)$ is smaller than that of $T_{cc}$. This is because the channel-coupling effect is suppressed for $X(3872)$ with the larger threshold energy difference in comparison with the $T_{cc}$ case. The compositeness of $T_{cc}$ is also discussed in Ref.~\cite{Albaladejo:2022sux} without channel couplings. They concluded the molecular dominance of $T_{cc}$, but also pointed out that the channel coupling may play an important role. In this work, we explicitly demonstrate how the channel coupling contributes to the compositeness.

To examine the cutoff dependence of the results, we perform the same analysis with cutoff $\Lambda=770$ MeV, in light of the $\rho$ meson exchange. The results are shown in Fig.~\ref{fig:apply-770}. Qualitatively, we find the same tendency as in Fig.~\ref{fig:apply}: strong coupled-channels (decay) effect for $T_{cc}$ [for $X(3872)$]. However, from quantitative comparison with Fig.~\ref{fig:apply}, we find that the coupled-channels effect is enhanced, the decay effect is suppressed, and the compositeness is increased for the larger cutoff. With $\Lambda=770$ MeV, the typical energy scales are $E_{\rm typ}=306$~MeV ($T_{cc}$) and $E_{\rm typ}=307$~MeV [$X(3872)$], and $\Delta\omega$, $\Gamma$, and $B$ are now regarded as small relative to $E_{\rm typ}$. As a consequence, the coupled-channels effect is more emphasized but the decay effect becomes less important. In particular, the decrease of $B/E_{\rm typ}$ lets the system be closer to the universality limit and hence the composite nature of the state becomes more prominent.

With the error bars in Figs.~\ref{fig:apply} and \ref{fig:apply-770}, we show the compositeness by taking into account the experimental errors of the eigenenergies. We find that the effect of the errors on the results of the compositenesses of $T_{cc}$ and $X(3872)$ is quantitatively small, as seen in the figures. The reason is understood as follows. The error of the binding energy of $T_{cc}$ in Eq.~\eqref{eq:Tcc-energy} is one order of magnitude smaller than the central value. As seen in Figs.~\ref{fig:apply} and \ref{fig:apply-770}, the compositeness of $T_{cc}$ is not affected by the decay width because the width is much smaller than the binding energy. In addition, the width of $T_{cc}$ in Eq.~\eqref{eq:Tcc-energy} includes the error which mainly contributes towards reducing the width. As a consequence, the compositeness of $T_{cc}$ does not change very much when we consider the errors of the mass and width. For $X(3872)$, the real part of the eigenenergy in Eq.~\eqref{eq:X3872-energy} can go above the threshold within the error. Nevertheless, the large imaginary part weakens its impact on the compositeness, because the errors change the magnitude of the complex eigenenergy only slightly. In summary, the experimental errors of the mass and width have only a minor effect on the compositeness, thanks to the small errors of $T_{cc}$ and the (relatively) large decay width of $X(3872)$. 

It is instructive to evaluate the probability of obtaining the model with the composite dominant state $P_{\rm comp}$ of $T_{cc}$ and $X(3872)$ discussed in the previous sections. $P_{\rm comp}$ is defined by Eq.~\eqref{eq:Pcomp} as the fraction of the parameter region where the state is composite dominant. Here, we examine two methods to determine $\nu_{c}$ in Eq.~\eqref{eq:Pcomp} for the different discussions. First, by focusing on the compositeness of the threshold channel $\tilde{X}_{1}$, we can discuss the low-energy universality as in Sec.~\ref{subsec:2ch}. In this case, we consider $P_{\rm comp}$ in terms of $\tilde{X}_{1}$ ($P_{\rm comp}^{\tilde{X}_{1}}$) with $\nu_{c}$ determined by the condition $\tilde{X}_{1}=0.5$. Second, because not only $\tilde{X}_{1}$ but also $\tilde{X}_{2}$ contributes to the molecular component, we can also determine $\nu_{c}$ by the condition $\tilde{X}_{1}+\tilde{X}_{2}=0.5$, and discuss $P^{\tilde{X}_{1}+\tilde{X}_{2}}_{\rm comp}$ to consider the molecular nature of $T_{cc}$ and $X(3872)$.

Let us evaluate $P_{\rm comp}$ of $T_{cc}$ and $X(3872)$. For $\Lambda=140$~MeV, we obtain
\begin{align}
P_{\rm comp}^{\tilde{X}_{1}}(T_{cc},\Lambda=140\ \rm{MeV})&=0.45^{+0.049}_{-0.037},\\
P_{\rm comp}^{\tilde{X}_{1}}[X(3872),\Lambda=140\ \rm{MeV}]&=0.59^{+0.040}_{-0.043}.
\end{align}
This result shows that the substantial coupled-channels and decay effects can reduce the threshold channel compositeness of $T_{cc}$ and $X(3872)$, even though both the states exist within the 1~MeV region from the threshold. The molecular component $P^{\tilde{X}_{1}+\tilde{X}_{2}}_{\rm comp}$ is calculated as follows:
\begin{align}
P^{\tilde{X}_{1}+\tilde{X}_{2}}_{\rm comp}(T_{cc},\Lambda=140\ \rm{MeV})&=0.71^{+0.012}_{-0.008},\\
P^{\tilde{X}_{1}+\tilde{X}_{2}}_{\rm comp}[X(3872),\Lambda=140\ \rm{MeV}]&=0.65^{+0.027}_
{-0.035}.
\end{align}
When the $\tilde{X}_{2}$ component is taken into account, the composite dominant region in the parameter space increases. For the cutoff $\Lambda=770$ MeV, we obtain the following results:
\begin{align}
P_{\rm comp}^{\tilde{X}_{1}}(T_{cc},\Lambda=770\ \rm{MeV})&=0.85^{+0.019}_{-0.009},\\
P_{\rm comp}^{\tilde{X}_{1}}[X(3872),\Lambda=770\ \rm{MeV}]&=0.87^{+0.016}_{-0.014}, \\
P_{\rm comp}^{\tilde{X}_{1}+\tilde{X}_{2}}(T_{cc},\Lambda=770\ \rm{MeV})&=0.94^{+0.004}_{-0.001},\\
P_{\rm comp}^{\tilde{X}_{1}+\tilde{X}_{2}}[X(3872),\Lambda=770\ \rm{MeV}]&=0.92^{+0.004}_{-0.008}.
\end{align}
As discussed above, the larger energy scale is introduced by the larger cutoff, and the composite nature of the bound state is more emphasized. In all cases, the errors of $P_{\rm comp}$ are small. This is because the compositeness is not affected by the experimental error, as discussed above.

The concrete value of $\nu_{0}$ cannot be determined in the effective field theory, unless other physical quantities (such as the scattering length) are given in addition to the eigenenergy. Alternatively, one can employ a specific model to estimate the value of $\nu_{0}$. For example, the constituent quark model in Ref.~\cite{Karliner:2017qjm} gives the bare energy of the four-quark state of $T_{cc}$ as $\nu_{0}=7$~MeV. From the estimated value of $\nu_{0}$, one can read off the structure of $T_{cc}$ and $X(3872)$ from Fig.~\ref{fig:apply}. 

In summary, the molecular nature of states can be modified by both the decay and the coupled-channels effects even if the pole exists near the threshold, and we need to consider these effects for the quantitative discussion of the compositeness. In particular, the coupled-channels effect should be important for $T_{cc}$, and decay width for $X(3872)$, as demonstrated in Figs.~\ref{fig:apply} and \ref{fig:apply-770}.


\section{Summary}
\label{sec:sum}

We have discussed the structure of weakly bound states from the viewpoint of the compositeness. Naively, the states near the two-body $s$-wave threshold are expected to have a molecule-type composite structure. In this paper, we explicitly demonstrate that it is always possible to construct a weakly bound noncomposite state, but only with a significant fine tuning in the system. In other words, the realization of the noncomposite state near the threshold is probabilistically suppressed, in accordance with the low-energy universality. In addition, we quantitatively study how this universal nature of the near-threshold bound states can be modified by the various effects, and examine their implications for the structure of exotic hadrons.

We first construct an effective field theory model with a bare state coupled to a single two-body scattering channel, and evaluate the compositeness of the bound state in this model within the allowed parameter region. With the assumption of naturalness, it is shown that the bound state in this model is usually an  elementary-dominant state originating from the bare state when the binding energy $B$ is of the order of the typical energy scale of the system, $B\sim E_{\rm typ}$. However, if the binding energy is small ($B\ll E_{\rm typ}$), then the bound state has a high probability of being the composite dominant state in the parameter region of the model. We quantitatively show that the probability of generating the composite dominant state gradually increases when $B$ decreases, and finally approaches unity in the $B\to 0$ limit. 

While this simple model captures the essential features of the near-threshold bound states, there are various effects present in the application to exotic hadrons. We thus generalize the above mentioned model by including the four-point contact interaction, the decay effect, and the coupling to the additional scattering channel. It is shown that the attractive (repulsive) four-point contact interaction increases (decreases) the compositeness of the bound state, because it helps to enhance (suppress) the generation of the molecule component. We show that the decay and coupled-channels effects decrease the compositeness, as they induce the contributions from the other components. The importance of the decay (coupled-channels) effect is characterized by the ratio of the decay width (the threshold energy difference) with respect to the real part of the eigenenergy. 

Finally, we consider the structure of $T_{cc}$ and $X(3872)$ in this perspective. It is known that $T_{cc}$ and $X(3872)$ appear close to the $D^{0}D^{*+}$ and $D^{0}\bar{D}^{*0}$ thresholds, respectively, but both states have a finite decay width and a nearby coupled channel [$D^{*0}D^{+}$ for $T_{cc}$ and $D^{+}D^{*-}$ for $X(3872)$]. As expected from the small threshold energy difference in $T_{cc}$ and the sizable decay width of $X(3872)$, we show that the channel coupling to $D^{*0}D^{+}$ (the decay effect) largely influences the compositeness of $T_{cc}$ [$X(3872)$]. In other words, it is important to consider the coupling to $D^{*0}D^{+}$ (the decay effect) to quantitatively study the internal structure of $T_{cc}$ [$X(3872)$]. In this way, we expect that the result of this work provides quantitative guidance to pin down the important effects for the discussion of the structure of near-threshold states.

\begin{acknowledgments}
The authors thank Atsushi Hosaka, Rich Lebed, Eulogio Oset, and Hagop Sazdjian for useful comments and discussions.
This work was supported in part by the Grants-in-Aid for Scientific Research from JSPS (Grants
No. JP23KJ1796, 
No. JP22K03637, and 
No. JP18H05402). 
This work was supported by JST, the establishment of university
fellowships towards the creation of science technology innovation, Grant No. JPMJFS2139.  
\end{acknowledgments}

\bibliography{refs.bib}

\end{document}